\documentclass[10pt,conference]{IEEEtran}

\usepackage[
  T1
]{fontenc}

\usepackage[
  nohyperlinks,
  nolist
]{acronym}

\usepackage[
  table
]{xcolor}

\usepackage{tcolorbox}

\usepackage{calc}

\usepackage{booktabs,tabularx,multicol,multirow,makecell}

\usepackage{caption,subcaption}

\usepackage{tikz,tikz-layers}
\usetikzlibrary{calc,positioning,arrows,automata,shapes.misc,shapes.geometric,decorations.markings}

\usepackage{xspace}

\usepackage{float}
\newfloat{codelisting}{htbp}{lop}
\floatname{codelisting}{List.}
\usepackage{url}

\usepackage{siunitx}
\DeclareSIUnit{\nothing}{\relax}

\usepackage{orcidlink}
\newcommand{\orcid}[1]{\space\orcidlink{#1}}

\usepackage{hyperref}

\usepackage{flushend}

\usepackage{listings}

\title{Temporal HAL-API Dependencies as a~Gateway to~Formal Embedded Software Development}
\author{
  \IEEEauthorblockN{Manuel~Bentele\orcid{0009-0003-4794-958X}}
  \IEEEauthorblockA{Hahn-Schickard~Institute\\Villingen-Schwenningen, Germany}
  \and
  \IEEEauthorblockN{Andreas~Podelski\orcid{0000-0003-2540-9489}}
  \IEEEauthorblockA{University~of~Freiburg\\Freiburg,~Germany}
  \and
  \IEEEauthorblockN{Axel~Sikora\orcid{0000-0003-0878-2919}}
  \IEEEauthorblockA{Hahn-Schickard~Institute\\Villingen-Schwenningen,~Germany}
  \and
  \IEEEauthorblockN{Bernd~Westphal\orcid{0000-0002-6824-0567}}
  \IEEEauthorblockA{German~Aerospace~Center\\Oldenburg,~Germany}
}

\hypersetup{
  pdftitle      ={Temporal HAL-API Dependencies as a Gateway to Formal Embedded Software Development},
  pdfauthor     ={Manuel~Bentele, Andreas~Podelski, Axel~Sikora, Bernd~Westphal},
  pdfkeywords   ={Hardware Abstraction Layer, Embedded Software, Formal Requirements, Static Analysis, Software Model Checking},
  pdfborderstyle={/S/U/W 1}
}

\input{macros.tex}
\colorlet{cgreen}{green}
\colorlet{cyellow}{yellow}
\colorlet{corange}{orange}
\colorlet{cred}{red}
\colorlet{cviolet}{violet}
\colorlet{cblue}{cyan}
\colorlet{cdlue}{blue}
\colorlet{cgray}{gray}
\colorlet{cblack}{black}
\colorlet{bgray}{gray!10}
\colorlet{lgray}{gray!5}
\newcommand{\bFillOpacity}{0.15}
\newcommand{\bFillIntensity}{15}

\newcommand*{\belowrulesepcolor}[1]{%
    \noalign{%
        \kern-\belowrulesep
        \begingroup
            \color{#1}%
            \hrule height\belowrulesep
        \endgroup
    }%
}
\newcommand*{\aboverulesepcolor}[1]{%
    \noalign{%
        \begingroup
            \color{#1}%
            \hrule height\aboverulesep
        \endgroup
        \kern-\aboverulesep
    }%
}

\makeatletter
\pgfkeys{%
  /tikz/on layer/.code={
    \def\tikz@path@do@at@end{\endpgfonlayer\endgroup\tikz@path@do@at@end}%
    \pgfonlayer{#1}\begingroup%
  }%
}
\makeatother
\tikzset{
    box/.style = {draw, rectangle, fill = none},
    font = \footnotesize,
    thadRoutine/.style={minimum width=13pt, minimum height=13pt},
    thadRoutineBox/.style={thadRoutine,draw, rectangle, fill=white,inner sep=3pt,text centered},
    thad/.style={preaction={on layer=background,line width=10pt,draw=cblue!\bFillIntensity},
    draw=cgray!50,
decoration = { markings, mark=at position #1 with {
\node[anchor=center,transform shape,inner sep=2.5pt, fill=cblue!\bFillIntensity] {$\triangleleft$}; }},
postaction=decorate },
    thad/.default=0.5,
    arrow/.style = {-triangle 45},
    crossNode/.style = {draw, fill = black, solid, circle, minimum size = 1mm, inner sep = 0pt},
    device/.style = {box, minimum width = 0.275\linewidth, minimum height = 1.5cm, inner sep = 0pt},
    master/.style = {fill = corange, fill opacity = 0.15},
    slave/.style = {fill = cgray, fill opacity = 0.15},
    pics/slaveFourInputsStyle/.style args = {#1/#2/#3/#4/#5}{
        code = {
            \node[device, slave, #5] (-sl) {};
            \node[anchor = east] at (-sl.east) {\tikzpictext};
            \coordinate (-p1) at ($(-sl.north west)!0.2!(-sl.south west)$);
            \coordinate (-p2) at ($(-sl.north west)!0.4!(-sl.south west)$);
            \coordinate (-p3) at ($(-sl.north west)!0.6!(-sl.south west)$);
            \coordinate (-p4) at ($(-sl.north west)!0.8!(-sl.south west)$);
            \node[font=\tiny,anchor=west] (-1) at (-p1) {#1};
            \node[font=\tiny,anchor=west] (-2) at (-p2) {#2};
            \node[font=\tiny,anchor=west] (-3) at (-p3) {#3};
            \node[font=\tiny,anchor=west] (-4) at (-p4) {#4};
        }
    },
    pics/masterFourInputs/.style args = {#1/#2/#3/#4}{
        code = {
            \node[device, master] (-ma) {};
            \node[anchor = west] at (-ma.west) {\tikzpictext};
            \coordinate (-p1) at ($(-ma.north east)!0.2!(-ma.south east)$);
            \coordinate (-p2) at ($(-ma.north east)!0.4!(-ma.south east)$);
            \coordinate (-p3) at ($(-ma.north east)!0.6!(-ma.south east)$);
            \coordinate (-p4) at ($(-ma.north east)!0.8!(-ma.south east)$);
            \node[font=\tiny,anchor=east] (-1) at (-p1) {#1};
            \node[font=\tiny,anchor=east] (-2) at (-p2) {#2};
            \node[font=\tiny,anchor=east] (-3) at (-p3) {#3};
            \node[font=\tiny,anchor=east] (-4) at (-p4) {#4};
        }
    },
    pics/masterSevenInputs/.style args = {#1/#2/#3/#4/#5/#6/#7}{
        code = {
            \node[device, master, minimum height = 2.4cm] (-ma) {};
            \node[anchor = west] at (-ma.west) {\tikzpictext};
            \coordinate (-p1) at ($(-ma.north east)!0.125!(-ma.south east)$);
            \coordinate (-p2) at ($(-ma.north east)!0.250!(-ma.south east)$);
            \coordinate (-p3) at ($(-ma.north east)!0.375!(-ma.south east)$);
            \coordinate (-p4) at ($(-ma.north east)!0.500!(-ma.south east)$);
            \coordinate (-p5) at ($(-ma.north east)!0.625!(-ma.south east)$);
            \coordinate (-p6) at ($(-ma.north east)!0.750!(-ma.south east)$);
            \coordinate (-p7) at ($(-ma.north east)!0.875!(-ma.south east)$);
            \node[font=\tiny,anchor=east] (-1) at (-p1) {#1};
            \node[font=\tiny,anchor=east] (-2) at (-p2) {#2};
            \node[font=\tiny,anchor=east] (-3) at (-p3) {#3};
            \node[font=\tiny,anchor=east] (-4) at (-p4) {#4};
            \node[font=\tiny,anchor=east] (-5) at (-p5) {#5};
            \node[font=\tiny,anchor=east] (-6) at (-p6) {#6};
            \node[font=\tiny,anchor=east] (-7) at (-p7) {#7};
        }
    },
    pics/swlapp/.style args = {#1}{
        code = {
            \node[minimum width=0.6\linewidth, minimum height=0.02\textheight] (-app) {#1};
            \coordinate (-c11) at ($(-app.north west) + (0, 0)$);
            \coordinate (-c12) at ($(-app.south west) + (0, -0.02\textheight)$);
            \coordinate (-c13) at ($(-app.south west) + (0.1\linewidth-3pt, -0.02\textheight)$);
            \coordinate (-c14) at ($(-app.south west) + (0.1\linewidth-3pt, 0)$);
            \coordinate (-c21) at ($(-app.north east) + (0, 0)$);
            \coordinate (-c22) at ($(-app.south east) + (0, -0.02\textheight)$);
            \coordinate (-c23) at ($(-app.south east) + (-0.1\linewidth+3pt, -0.02\textheight)$);
            \coordinate (-c24) at ($(-app.south east) + (-0.1\linewidth+3pt, 0)$);
            \begin{pgfonlayer}{behind}
                \draw[fill = cred,fill opacity=\bFillOpacity] (-c11) -- (-c12) -- (-c13) -- (-c14) -- (-c24) -- (-c23) -- (-c22) -- (-c21) -- cycle;
            \end{pgfonlayer}
        }
    },
    pics/swlapp/.default=Application,
    pics/swlint/.style args = {#1/#2/#3}{
        code = {
            \node[minimum width=0.6\linewidth, minimum height=0.06\textheight] (-oshal) {};
            \coordinate (-c10) at ($(-oshal.south west) + (0,0)$);
            \coordinate (-c11) at ($(-oshal.south west) + (0,0.02\textheight)$);
            \coordinate (-c12) at ($(-oshal.north west) + (0,-0.02\textheight)$);
            \coordinate (-c13) at ($(-oshal.north west) + (0.1\linewidth,-0.02\textheight)$);
            \coordinate (-c14) at ($(-oshal.north west) + (0.1\linewidth,0)$);
            \coordinate (-c20) at ($(-oshal.south east) + (0,0)$);
            \coordinate (-c21) at ($(-oshal.south east) + (0,0.02\textheight)$);
            \coordinate (-c22) at ($(-oshal.north east) + (0,-0.02\textheight)$);
            \coordinate (-c23) at ($(-oshal.north east) + (-0.1\linewidth,-0.02\textheight)$);
            \coordinate (-c24) at ($(-oshal.north east) + (-0.1\linewidth,0)$);
            \draw (-c10) -- (-c11) -- (-c12) -- (-c13) -- (-c14) -- (-c24) -- (-c23) -- (-c22) -- (-c21) -- (-c20) -- cycle;
            \draw[dashed] (-c11) -- (-c21);
            \draw[dashed] (-c13) -- (-c23);
            \begin{pgfonlayer}{behind}
                \fill[fill = cblue, fill opacity=\bFillOpacity] (-c11) -- (-c12) -- (-c22) -- (-c21);
                \fill[fill = cgreen, fill opacity=\bFillOpacity] (-c13) -- (-c14) -- (-c24) -- (-c23);
                \fill[fill = corange, fill opacity=\bFillOpacity] (-c10) -- (-c11) -- (-c21) -- (-c20);
            \end{pgfonlayer}
            \node[anchor=center] at ($(-oshal.south) + (0,0.05\textheight)$) {#1};
            \node[anchor=center] at ($(-oshal.south) + (0,0.03\textheight)$) {#2};
            \node[anchor=center] at ($(-oshal.south) + (0,0.01\textheight)$) {#3};
        }
    },
    pics/swlint/.default=\acs{hal}-\acs{api}/\acs{hal}~implementation/\acs{spi}~subsystem,
    pics/swlintst/.style args = {#1/#2}{
        code = {
            \node[minimum width=0.6\linewidth, minimum height=0.04\textheight] (-midhal) {};
            \coordinate (-c100) at ($(-midhal.south west) + (0.1\linewidth-3pt,0)$);
            \coordinate (-c10) at ($(-midhal.south west) + (0.1\linewidth-3pt,-0.02\textheight)$);
            \coordinate (-c11) at ($(-midhal.south west) + (0,-0.02\textheight)$);
            \coordinate (-c12) at ($(-midhal.north west) + (0,-0.02\textheight)$);
            \coordinate (-c13) at ($(-midhal.north west) + (0.1\linewidth,-0.02\textheight)$);
            \coordinate (-c14) at ($(-midhal.north west) + (0.1\linewidth,0)$);
            \coordinate (-c200) at ($(-midhal.south east) + (-0.1\linewidth+3pt,0)$);
            \coordinate (-c20) at ($(-midhal.south east) + (-0.1\linewidth+3pt,-0.02\textheight)$);
            \coordinate (-c21) at ($(-midhal.south east) + (0,-0.02\textheight)$);
            \coordinate (-c22) at ($(-midhal.north east) + (0,-0.02\textheight)$);
            \coordinate (-c23) at ($(-midhal.north east) + (-0.1\linewidth,-0.02\textheight)$);
            \coordinate (-c24) at ($(-midhal.north east) + (-0.1\linewidth,0)$);
            \draw (-c100) -- (-c10) -- (-c11) -- (-c12) -- (-c13) -- (-c14) -- (-c24) -- (-c23) -- (-c22) -- (-c21) -- (-c20) -- (-c200) -- cycle;
            \draw[dashed] (-c13) -- (-c23);
            \begin{pgfonlayer}{behind}
                \fill[fill = cblue, fill opacity=\bFillOpacity] (-c100) -- (-c10) -- (-c11) -- (-c12) -- (-c22) -- (-c21) -- (-c20) -- (-c200);
                \fill[fill = cgreen, fill opacity=\bFillOpacity] (-c13) -- (-c14) -- (-c24) -- (-c23);
            \end{pgfonlayer}
            \node[anchor=center] at ($(-midhal.south) + (0,0.03\textheight)$) {#1};
            \node[anchor=center] at ($(-midhal.south) + (0,0.01\textheight)$) {#2};
        }
    },
    pics/swlintst/.default=\acs{hal}-\acs{api}/\acs{hal}~implementation,
    pics/swlintend/.style args = {#1/#2}{
        code = {
            \node[minimum width=0.6\linewidth, minimum height=0.04\textheight] (-endhal) {};
            \coordinate (-c11) at ($(-endhal.south west) + (0,0)$);
            \coordinate (-c12) at ($(-endhal.north west) + (0,-0.02\textheight)$);
            \coordinate (-c13) at ($(-endhal.north west) + (0.05\textwidth,-0.02\textheight)$);
            \coordinate (-c14) at ($(-endhal.north west) + (0.05\textwidth,0)$);
            \coordinate (-c21) at ($(-endhal.south east) + (0,0)$);
            \coordinate (-c22) at ($(-endhal.north east) + (0,-0.02\textheight)$);
            \coordinate (-c23) at ($(-endhal.north east) + (-0.05\textwidth,-0.02\textheight)$);
            \coordinate (-c24) at ($(-endhal.north east) + (-0.05\textwidth,0)$);
            \draw (-c11) -- (-c12) -- (-c13) -- (-c14) -- (-c24) -- (-c23) -- (-c22) -- (-c21) -- cycle;
            \draw[dashed] (-c13) -- (-c23);
            \begin{pgfonlayer}{behind}
                \fill[fill = cblue, fill opacity=\bFillOpacity] (-c11) -- (-c12) -- (-c22) -- (-c21);
                \fill[fill = cgreen, fill opacity=\bFillOpacity] (-c13) -- (-c14) -- (-c24) -- (-c23);
            \end{pgfonlayer}
            \node[anchor=center] at ($(-endhal.south) + (0,0.03\textheight)$) {#1};
            \node[anchor=center] at ($(-endhal.south) + (0,0.01\textheight)$) {#2};
        }
    },
    pics/swlintst/.default=\acs{hal}-\acs{api}/\acs{hal}~implementation,
    pics/swlhw/.style args = {#1}{
        code = {
            \node[draw,fill=cgray,fill opacity=\bFillOpacity, text opacity=1, minimum width=0.6\linewidth, minimum height = 0.02\textheight] (-hw) {#1};
        }
    },
    pics/swlhw/.default=Hardware
}

\begin{document}

\maketitle

\begin{acronym}
  \acro{hardware}[HW]{Hardware}
  \acro{software}[SW]{Software}
  \acro{hal}[HAL]{Hardware Abstraction Layer}
  \acro{api}[API]{Application Programming Interface}
  \acro{thad}[THAD]{Temporal \acs{hal}-\acs{api} Dependency}
  \acroplural{thad}[THADs]{Temporal \acs{hal}-\acs{api} Dependencies}
  \acro{misra}[MISRA]{Motor Industry Software Reliability Association}
  \acro{ast}[AST]{Abstract Syntax Tree}
  \acro{pin}[PIN]{Personal Identification Number}
  \acro{smt}[SMT]{Satisfiability Modulo Theories}
  \acro{usb}[USB]{Universal Serial Bus}
  \acro{spi}[SPI]{Serial Peripheral Interface}
  \acro{cpu}[CPU]{Central Processing Unit}
  \acro{sclk}[SCLK]{Serial Clock}
  \acro{mosi}[MOSI]{Master Out/Slave In}
  \acro{miso}[MISO]{Master In/Slave Out}
  \acro{ss}[SS]{Slave Select}
  \acro{lsb}[LSB]{Least Significant Bit}
  \acro{lsbfe}[LSBFE]{\acs{lsb}-First Enable}
  \acro{cpol}[CPOL]{Clock Polarity}
  \acro{cpha}[CPHA]{Clock Phase}
  \acro{posix}[POSIX]{Portable Operating System Interface}
  \acro{vfs}[VFS]{Virtual File System}
  \acro{acsl}[ACSL]{ANSI/ISO C~Specification Language}
  \acro{svcomp}[SV-COMP]{Competition on Software Verification}
  \acro{loc}[LOC]{Lines of Code}
  \acro{cegar}[CEGAR]{Counter-Example Guided Abstraction Refinement}
  %-----
  \acused{hardware}
  \acused{software}
  \acused{misra}
  \acused{pin}
  \acused{usb}
  \acused{smt}
  \acused{cpu}
  \acused{sclk}
  \acused{mosi}
  \acused{miso}
  \acused{ss}
  \acused{lsb}
  \acused{lsbfe}
  \acused{cpol}
  \acused{cpha}
  \acused{posix}
  \acused{loc}
\end{acronym}

\begin{abstract}
  \acp{thad} can be useful to capture an interesting class of correctness properties in embedded software development.
  They demand a moderate effort for specification (which can be done via program annotations) and verification (which can be done automatically via software model checking).
  In this sense, they have the potential to form an interesting sweet spot between generic properties (that demand virtually no specification effort, and that are typically addressed by static analysis) and application-specific properties as addressed by full-fledged formal methods.
  Thus, they may form a gateway to wider and more economic use of formal methods in industrial embedded software development.
\end{abstract}

\section{Introduction}
\label{sec:introduction}
As a possible research strategy for the investigation of formal methods, as well as for an early investigation of principles, it seems useful to single out a practically relevant class of correctness properties that demand a moderate effort for specification and that can be checked automatically.
We present such a class of correctness properties, for a specific class of programs, together with a preliminary case study.

The developer can formally specify the correctness properties using \emph{\acfp{thad}}, a class of formulas which we introduce (with a formal syntax and semantics).
Alternatively, the developer can formally specify the correctness properties by a \emph{program annotation} (which uses the existing syntax and semantics of program statements).
The verification of \acp{thad} can be done automatically via software model checking.
Thus, in terms of the burden put on the developer, the corresponding formal method lies minimally above static analysis methods that are used to check \emph{generic} correctness properties (such as, \eg, the properties referring to array bounds).
Thus, they can be seen as a possible next step towards formal methods for more involved properties expressing, \eg, aspects of the functional correctness of the underlying application.

By formulating a set of examples of correctness properties for a specific class of programs, the case study provides benchmarks which may be useful beyond the work presented here, and in particular for the investigation of alternative or competing approaches.

In the remainder of this section, we will discuss \acp{thad}, first in the general context of embedded software development, and then in the context of two motivational examples.

\subsection{\acsp{thad} for embedded software}
\label{sec:introduction:thads}
In the context of embedded software development, the practical importance of checking \acp{thad} cannot be overestimated.
A \ac{hal} is composed of an implementation and an interface providing the routines which the upper software layer (the application program) can use to access the hardware.
The idea is that the developer can write hardware-independent programs and preserve portability across different hardware.
In principle, the documentation should not only specify the routines that can be called but also the temporal dependencies between them.
For example, the \emph{send} functionality of a bus provided by the routine \halApiRoutineSendS (which transmits data packets over the bus) may depend on a previous initialization of the hardware through the routine \halApiRoutineInitS.
The violation of the specification of such a dependency can lead to fatal errors, perhaps even hardware damage or hardware destruction.
It is notoriously difficult to track or analyze the corresponding error.

The correct usage of interfaces has been recognized as an important problem in general (\qty{18.6}{\percent} of all errors~\cite{Ko2003}, more than \qty{50}{\percent} of bug reports~\cite{Monperrus2013}), but the issue is exacerbated in the context of hardware dependent software.
A study reports that \qty{78}{\percent} of the bugs found are \ac{hardware}/\ac{software} interface bugs~\cite{Youssef2004}.

A lot of research, including ours~\cite{Nsiah2018}, \cite{Rathfelder2015}, has gone into frameworks that help separate  generic communication functionalities and application-specific functionalities.
Frameworks with high-level interface description languages, for example, can help developers to specify and implement the communication interfaces for sensor networks; see, \eg, \cite{Nsiah2018}, \cite{Rathfelder2015}.

\acp{hal} can generally help the developer in mastering complexity~\cite{Popovici2009}, \cite{Yoo2003}.
In a way, however, they shift the problem upwards, \ie, from the hardware interface to the software using it.
The verification of the application program is known to be difficult; it has been found to take up at least one third of the overall development cost~\cite{Lettnin2017}.
Recent investigations confirm that, even with comprehensive testing, bugs can remain undetected for years \cite{Eichelberger2017}.
Dynamic verification, from unit testing to system testing, remains the common approach~\cite{AbbaspourAsadollah2015}.
The use of static analysis is still very limited in this context~\cite{Anderson2008}.
A notable exception here is the application of the \acs{misra}-C Checker for checking the conformance with coding standards~\cite{Bagnara2018}.

Although failures due to (possibly rare) sequences of \ac{hal} routine calls can have nasty consequences, the topic of methods to formalize the corresponding error properties and guarantee the absence of such failures remains under-explored.

In contrast, the verification of temporal interface specifications for Windows or Linux \emph{device drivers} has received a lot of attention; in fact, it has been one of the driving motivations of software model checking, and part of its success story; see, for example, Static Driver Verifier, which became a part of Visual Studio~\cite{Ball2006}.
One new challenge that comes with our focus on embedded software lies in the fact that the developer needs to define the temporal interface specifications anew for each \ac{hal}.
This may be a non-trivial task for the developers even if they know the hardware requirements and the resulting dependencies for the \ac{hal}.
The difficulty is perhaps exacerbated if the specification requires a new logic (with formal syntax and semantics).

For the purpose of comparison, it may be interesting to analyze the origin of the incentive for writing the specification.
In the case of device drivers for personal computers, the provider of the operating system (say, Microsoft) defines temporal interface specifications for device drivers as a protective measure (to cite from \cite{Ball2006}: bugs in kernel-level device drivers caused \qty{85}{\percent} of the system crashes in the Windows~XP operating system).
In the case of an embedded system, the developer may want a certificate which can be used to deflect blame in the event of a fault (\ie, a proof that the error does not lie in the application program but in the implementation of the \ac{hal} or elsewhere).

The term \emph{software model checking} refers to a new approach to program analysis integrated with automatic abstraction; the automatic abstraction here follows the scheme of \emph{\ac{cegar}}~\cite{Clarke2000}.
A handbook article \cite{Jhala2018} gives an introduction to the underlying principles, which build on a rich body of research on mathematical structures (fixpoints, Galois connections, Craig interpolants, \etc) and algorithms (fixpoint solving, constraint solving, \acs{smt} solving, decision procedures, \etc).

The annual international \ac{svcomp} witnesses to the maturity of software model checking; see, \eg, \cite{Beyer2017} for a systematic comparison of the participants including our own software model checker \ultimateAutomizer~\cite{Heizmann2018}.\footnote{\ultimateAutomizer is part of the open-source \ultimate program analysis framework, available at \url{https://ultimate-pa.org}.}

As for its place within the software development process,  a software model checker is often viewed as a digital assistant to the software developer, \ie, a tool that can answer questions about the behavior of a program upon ``push-button''.
It is thus viewed as a glorified testing tool (one which guarantees to find all errors and which can prove the absence of errors).
In contrast with testing, however, software model checking has not yet found its place in mainstream software development.

We here concentrate on formal methods based on software model checking, but our work prepares the ground for exploring alternative approaches in this context (and comparing them, \eg, on the benchmarks provided by our case study).
We here think, for example, of behavioral type refinements, which can be represented by typestates; they have mostly been investigated in the area of programming languages, in the context of type checking and type inference; see, \eg, \cite{Strom1986} and the rich body of ensuing research.
Another related line of work in this context is the detection of \ac{api} misuses; see, \eg, \cite{Amann2016} and~\cite{Amann2019} where abstract syntax trees are used to represent \ac{api} (mis)uses.

Approaches for the inference of (safe and permissive) interfaces for software libraries can be based on software model checking; see, \eg,~\cite{Henzinger2005}.
These approaches may be interesting in particular for more pervasive verification efforts including the correctness of lower layers, \eg, the \ac{hal} implementation.

\subsection{Motivational examples}
\label{sec:introduction:examples}

\begin{figure}[b]
  \centering
  \setlength{\fboxsep}{0pt}%
  \setlength{\fboxrule}{0.5pt}%
  \begin{subfigure}[t]{2.128cm}
    \centering
    \includegraphics[height=4cm,keepaspectratio]{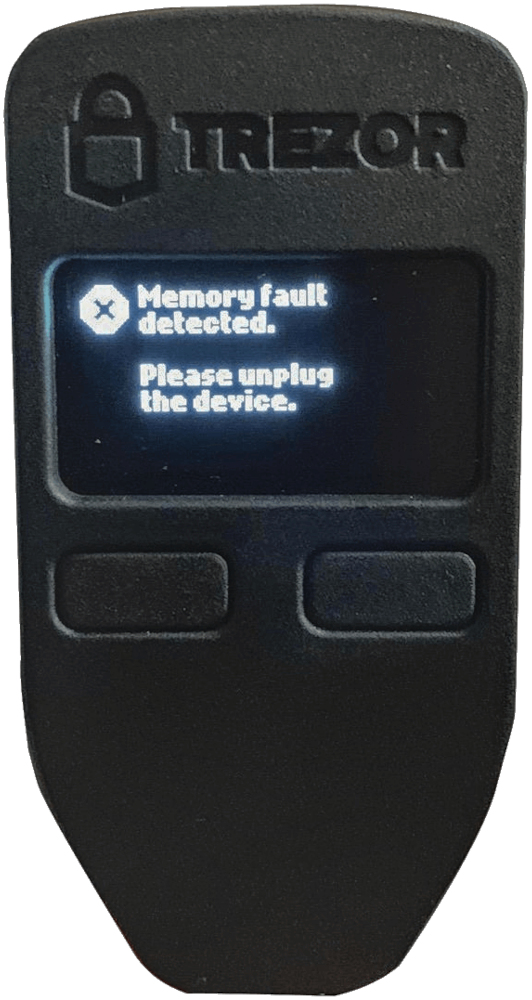}%
    \subcaption{\trezor hardware wallet.}
    \label{fig:trezor}
  \end{subfigure}
  \hfil
  \begin{subfigure}[t]{4.8cm+1pt}
    \centering
    \fbox{%
      \includegraphics[height=4cm,keepaspectratio]{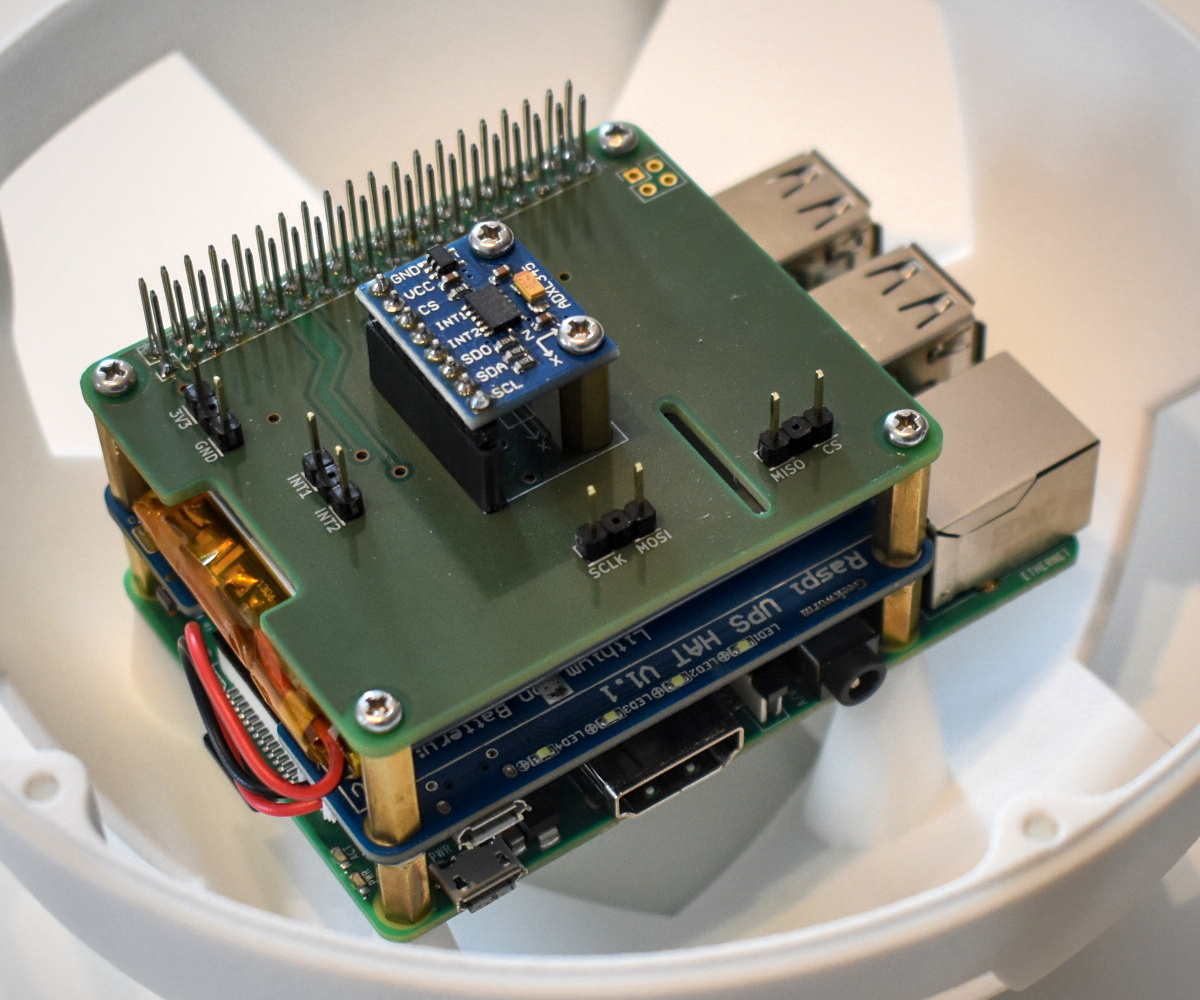}%
    }
    \subcaption{ADXL345 acceleration sensor as part of an embedded system.}
    \label{fig:demonstrator}
  \end{subfigure}
  \caption{Two example systems to illustrate the motivation for checking temporal \ac{hal} interface specifications.}
  \label{fig:examples}
\end{figure}

We can use a bug in a recent version of the \trezor hardware wallet%
\footnote{Bug in the \trezor hardware wallet reported at \url{https://www.reddit.com/r/TREZOR/comments/b983b1/solved_memory_fault_detected/}}
to illustrate the motivation for checking the correctness \wrt temporal \ac{hal} interface specifications (see \autoref{fig:trezor}).
The bug leads to a crash (memory fault) during the firmware update in the rather rare situation where the last action before the update was that the user entered a wrong personal four-digit \ac{pin} without entering the right \ac{pin} afterwards.
There is a persistent \ac{pin} failure counter that counts the number of consecutive attempts to enter a \ac{pin}.
This counter is reset to zero after the \ac{pin} is entered correctly.
When unlocking the storage, the code checks the counter.
If the counter is non-zero, the code enters a sleep phase, whose duration depends on the counter; in normal operations this code is used to prevent too many \ac{pin} tries, but during firmware upgrade the same code is called to unlock the storage to perform an upgrade to the next firmware version.
The reason for the crash is that a common sleep routine is used, which checks (or polls) the \ac{usb} hardware for incoming packets, but at this point in the firmware update, the \ac{usb} had not yet been initialized.
That is, the routines of the \ac{hal} are called in the wrong sequence; the \ac{usb} should not be used before it has been initialized.
A \ac{thad} will rule out this kind of behavior by prescribing the possible temporal ordering in sequences of \ac{hal} routine calls, such as \code{initialize()} and \code{poll()}.

Effects of \ac{thad} violations need not be as evident as the \trezor bug above as the following example shows.
Single-board computers such as the \emph{Raspberry~Pi} enjoy some popularity as a prototyping platform for embedded systems.
To develop embedded applications that need data from a digital acceleration sensor one would, for example, connect an \emph{ADXL345} sensor to the computer via the builtin, low-cost short-distance serial bus \acs{spi} (see \autoref{fig:demonstrator}, ADXL345 in blue on top).
Writing data to the bus configures the sensor, reading data from the bus yields acceleration data.
By properties of the \acs{spi} bus, a set of \acp{thad} would require that a set of routines are called that ensure that the controller is properly configured (transmission speed, encoding, \etc) because there is no standardized default value.
Violating one \ac{thad} by, \eg, forgetting one of the configuration routines could lead to particularly hard to spot malfunction of the software:
Reading data would succeed, if the unset configuration parameter has a suitable default.
If the default changes with a different hardware version, or depends on on previous uses of the bus after a warm-start, then the device would appear as having spurious data outages.
Errors of this kind are very hard to detect with conventional software engineering methods, in particular in the domain of embedded software where development and target platform are different and the target platform is often limited in its means for error analysis.

As illustrated in both examples, when a correctness property is not central to the definition or purpose of an embedded system, it tends to receive less attention during design.
Such a property is often expected to hold by default and, in fact, it does hold in the most common situations, then this is perhaps when its validation calls for a formal method (\ie, a method that finds all bugs resp.\@ gives a guarantee if there aren't any).

\section{\aclp{thad} on the Example of the \acl{spi}}
\label{sec:spi}
In this section, we introduce the concept and the relevance of \acp{thad} on the example of an \ac{api} for the \ac{spi}, which is a versatile, low-cost bus-system for short distance communication that is widely used in embedded systems.
The \ac{spi} case study is challenging (and to a certain extent representative for the embedded domain) since the protocol supports different hardware configurations and a multitude of hardware vendors offers \ac{spi} controllers with slightly different programming models, hence there is a need for a \ac{hal}.
One reason why we present this section in full technical detail is to prepare the grounds for follow-up work which investigates (and compares) the potential of a whole set of alternative approaches to check and possibly infer (different formalizations of) dependencies (approaches based, \eg, on typestates, as mentioned above).

\subsection{\acl{spi}: \acl{hardware}}
\label{sec:spi:hw}
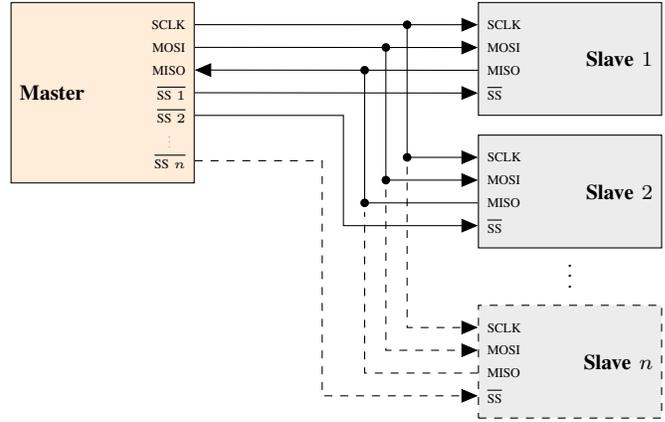
\begin{figure}
  \centering
  \begin{tikzpicture}
    \pic (ma) [pic text = {\textbf{Master}}] {masterSevenInputs = \acs{sclk}/\acs{mosi}/\acs{miso}/\textoverline{\acs{ss}~$1$}/\textoverline{\acs{ss}~$2$}/\scalebox{0.5}{\vdots}\hspace*{2mm}/\textoverline{\acs{ss}~$n$}};
    \pic (s1) [pic text = {\textbf{Slave~$1$}}, right = 0.425\linewidth of ma-ma.north east, anchor = north west] {slaveFourInputsStyle = \acs{sclk}/\acs{mosi}/\acs{miso}/\textoverline{\acs{ss}}/};
    \pic (s2) [pic text = {\textbf{Slave~$2$}}, below = 0.25 of s1-sl] {slaveFourInputsStyle = \acs{sclk}/\acs{mosi}/\acs{miso}/\textoverline{\acs{ss}}/};
    \pic (sn) [pic text = {\textbf{Slave~$n$}}, below = 0.75 of s2-sl] {slaveFourInputsStyle = \acs{sclk}/\acs{mosi}/\acs{miso}/\textoverline{\acs{ss}}/dashed};

    \node[inner sep = 0pt] at ($(s2-sl)!0.45!(sn-sl)$) {\vdots};

    \draw [arrow] (ma-p1) -- (s1-p1);
    \draw [arrow] (ma-p2) -- (s1-p2);
    \draw [arrow] (s1-p3) -- (ma-p3);
    \draw [arrow] (ma-p4) -- (s1-p4);

    \draw [arrow] ($(ma-p1)!0.750!(s1-p1)$) node[crossNode]{} |- node(e1) {} (s2-p1);
    \draw [arrow] ($(ma-p2)!0.675!(s1-p2)$) node[crossNode]{} |- node(e2) {} (s2-p2);
    \draw [arrow,-] (s2-p3) -| node(e3) {} ($(ma-p3)!0.600!(s1-p3)$) node[crossNode]{};
    \draw [arrow] (ma-p5) -| ($(ma-p5 |- s2-p4)!0.525!(s2-p4)$) |- node(e4) {} (s2-p4);

    \draw [arrow,dashed] (e1) node[crossNode]{} |- (sn-p1);
    \draw [arrow,dashed] (e2) node[crossNode]{} |- (sn-p2);
    \draw [arrow,-,dashed] (sn-p3) -| (e3) node[crossNode]{};
    \draw [arrow,dashed] (ma-p7)  -| ($(ma-p7 |- sn-p4)!0.450!(sn-p4)$) |- (sn-p4);
\end{tikzpicture}
  \caption{Example \ac{spi} hardware topology: Full-duplex star.}
  \label{fig:spi:star-topology}
\end{figure}

\ac{spi}~\cite{Hill1990} is a serial communication interface that is used to exchange data between hardware components of an embedded system.
If an embedded system needs, for example, acceleration data, the system may use a dedicated accelerometer chip, which includes an \ac{spi} controller, and wire this component to the embedded \ac{cpu}, which also includes an \ac{spi} controller.
\autoref{fig:spi:star-topology} shows a schematical wiring between \ac{spi} controllers.
The top-left node `Master' could, in the example, be the \ac{spi} controller in the \ac{cpu} and the top-right node `Slave~$1$' the \ac{spi} controller in the accelerometer peripheral.
Nodes exchange data over the \ac{ss} lines by a proper operation of the \ac{sclk}, \ac{mosi}, and \ac{miso} lines.

The \ac{spi} interconnect is attractive for embedded system designs due to its low complexity and cost (compared to, \eg, the \ac{usb} interconnect mentioned in the introduction), its availability (\ac{spi} controllers are integrated into many hardware components such as computer processors, microcontrollers, and digital sensors or actuators), and its flexibility (different topologies are supported, different data transmission modes, different transmission speeds, \etc).

\begin{table}[b]
  \centering
  \caption{\acs{spi} configuration parameters.}
  \label{tab:spi:configuration-parameters}
  \begin{tabularx}{\linewidth}{>{\hsize=.76\hsize}X>{\hsize=1.52\hsize}X>{\hsize=.72\hsize}X}
    \toprule
    \belowrulesepcolor{bgray}
    \rowcolor{bgray}\textbf{Parameter} & \textbf{Description}                     & \textbf{Typical~values} \\
    \aboverulesepcolor{bgray}
    \midrule
    \acs{cpol}                         & Polarity of \acs{sclk} during idle state & \num{0}, \num{1} \\
    \acs{cpha}                         & Phase of \acs{sclk} for data sampling    & \num{0}, \num{1} \\
    \acs{lsbfe}                        & Data transfer bit order of a word        & \num{0}, \num{1} \\
    Word~Size                          & Data transfer word size (in bit)         & \num{8}, \num{16}, \dots \\
    \acs{sclk}~frequency               & Data transfer frequency (in Hz)          & \qty{500}{\kilo\nothing}, \qty{1}{\mega\nothing}, \dots \\
    \bottomrule
  \end{tabularx}
\end{table}

\autoref{tab:spi:configuration-parameters} shows the most common parameters of the \ac{spi} protocol as supported by the \ac{spi} controller manufacturer Motorola~\cite{Motorola2003}.
Next to canonical communication protocol parameters such as the transmission speed (\ac{sclk}~frequency), the number of bits in a data unit (Word~Size), and the bit numbering \ac{lsb} (\ac{lsbfe}, right- or left-most bit of a word is transmitted first), the \ac{spi} protocol can also be configured for different polarities of the synchronous clock signal on the \ac{sclk} line (\ac{cpol}), and whether the \ac{ss} line is sampled on rising or falling edges of the \ac{sclk} line (\ac{cpha}).

There is a broad compatibility between \ac{spi} controllers from different manufacturers on the wiring side (as shown in \autoref{fig:spi:star-topology}), that is, if all \ac{spi} controllers in `Master' and `Slave~$1$' to `Slave~$n$' are configured to the same data transmission mode and speed, data exchange is simple and reliable.
Yet there is no universally agreement on default configuration for \ac{spi} controllers, and, even worse, the way how parameters are set programmatically (that is, which value needs to be written to which register in order to configure positive polarity of the bus clock \ac{sclk}) is not standardized.
Although most manufacturers of \ac{spi} peripherals adhere closely to the Motorola document~\cite{Motorola2003} as a de-facto standard, there are slight differences between \ac{spi} controllers offered by different manufacturers and hence there is a strong need for a \ac{spi} to ease the work of embedded programmers.

\subsection{Operating \acs{spi} by the \acl{hal} \spidev}
\label{sec:spi:hal}
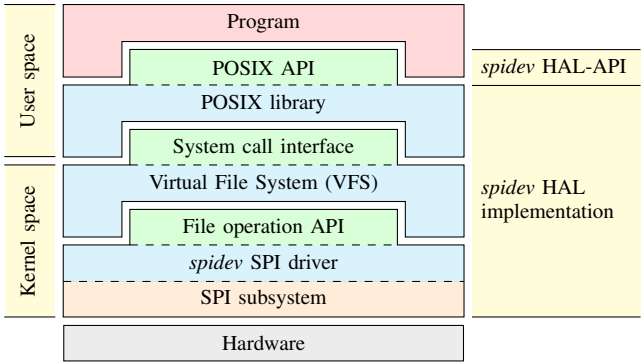
\begin{figure}
  \centering
  \begin{tikzpicture}[
        node distance = 3pt,
        grpDesc/.style = {fill = cyellow, fill opacity=\bFillOpacity, text opacity=1, minimum height = 0.075\linewidth},
        halDesc/.style = {align=left, minimum width = 0.225\linewidth,text width=0.225\linewidth }
    ]
    \pic (app) {swlapp=Program};
    \pic [below = of app-app] (libc) {swlintst=POSIX~API/POSIX~library};
    \pic [below = of libc-midhal] (vfs) {swlintst=System~call~interface/\ac{vfs}};
    \pic [below = of vfs-midhal] (spidevimpl) {swlint=File~operation~API/\spidev~SPI~driver/SPI subsystem};
    \pic [below = of spidevimpl-oshal] (hw) {swlhw};
    \node [rotate=90, grpDesc, left = of app-app.north west, anchor=south east, minimum width = 0.08\textheight+3pt+\pgflinewidth] (us) {User~space};
    \draw (us.north west) -- (us.south west);
    \draw (us.north east) -- (us.south east);
    \node [rotate=90, grpDesc, left = of spidevimpl-oshal.south west, anchor=south west, minimum width = 0.08\textheight+3pt+\pgflinewidth] (ks) {Kernel~space};
    \draw (ks.north west) -- (ks.south west);
    \draw (ks.north east) -- (ks.south east);
    \coordinate (chelp) at ($(libc-midhal.south east) + (0,0.03\textheight)$);
    \node [halDesc, right = of chelp, anchor=west, minimum height=0.02\textheight, halDesc] (halApi) {\spidev~HAL-API};
    \draw (halApi.north west) -- (halApi.north east);
    \node [halDesc, right = of spidevimpl-oshal.south east, anchor=south west, minimum height=0.12\textheight+6pt+1.5\pgflinewidth] (halImpl) {\spidev \acs{hal}\\implementation};
    \draw (halImpl.north west) -- (halImpl.north east);
    \draw (halImpl.south west) -- (halImpl.south east);
    \begin{pgfonlayer}{behind}
        \fill[cyellow, opacity = \bFillOpacity] (halApi.north west) rectangle (halImpl.north east);
        \fill[cyellow, opacity = \bFillOpacity] (halImpl.north west) rectangle (halImpl.south east);
    \end{pgfonlayer}
\end{tikzpicture}
  \caption{Overview of software layers from the \spidev~\acs{hal}.}
  \label{fig:spi:hal-spidev-layers}
\end{figure}

\autoref{fig:spi:hal-spidev-layers} illustrates the use of \ac{spi} buses from the perspective of the embedded systems programmer, who develops an embedded application (`Program' in \autoref{fig:spi:hal-spidev-layers}) which is supposed to process acceleration data that is provided by a dedicated accelerometer with integrated \ac{spi} controller.
The accelerometer is then part of the `Hardware' layer in \autoref{fig:spi:hal-spidev-layers}.
The (software) layers between `Hardware' and `Program' provide a \ac{hal} that is supposed to offer to the embedded systems programmer an interface that allows him or her to program on the level of \emph{general concepts} of \ac{spi} without the need to worry about \ac{spi} implementation details such as the memory mapping and the particular encoding of semantical parameter values such as `left-most bit transmitted first'.

A prominent example of a \ac{hal} for (in particular) \ac{spi} controllers is the Linux kernel.
The \ac{spi} subsystem and the \spidev driver (\cf bottom of \autoref{fig:spi:hal-spidev-layers}) encapsulate the implementation details to support \ac{spi} controllers from different manufacturers.
The layers on top of these are specific to the Linux operating system architecture where any kind of data input and output is abstracted into the concept of a file.
That is, to read data from the accelerometer in the example, the program uses the well-known routine \halApiRoutineOpenS to create a file descriptor for a device file from the system file system.
Reading and writing data from and to the \ac{spi} bus then amounts to calling \halApiRoutineReadS and \halApiRoutineWriteS as one would do for regular files on the system's storage devices (like hard disks).
Technically, the abstraction to file access is provided by the \ac{vfs} which is used by the \ac{posix} layer that provides routines by which in particular the \ac{spi} peripherals can be operated.

The \ac{api} of the \ac{spi} \ac{hal} in the Linux kernel is summarized by \autoref{tab:hal:spidev-api}.
There is \halApiRoutineOpenS, which returns a file descriptor~$\mathit{fd}$ (or an error code) when called for a device file that corresponds to an \ac{spi} controller.
Given a valid file descriptor~$\mathit{fd}$, \halApiRoutineReadS, \halApiRoutineWriteS, and \halApiRoutineCloseS can be used as usual.
Further operations on devices are accessed through the general purpose routine \halApiRoutineIoctlS:
The first parameter is always a file descriptor, the second parameter selects the function, which may consider the values of further parameters.
In the \ac{spi} \acs{hal}-\acs{api}, full-duplex transfer is available via \halApiRoutineIoctlS (routines \halApiRoutineReadS and \halApiRoutineWriteS only support half-duplex transfer).
More prominently, setting and getting the \ac{spi} parameters listed in \autoref{tab:spi:configuration-parameters} is done through the \halApiRoutineIoctlS routine.
For example, to set the bit order to `left-most bit transmitted first', the application program would include a call like \halApiRoutineIoctlWLSBFFD where \halApiRoutineParamValueWLSBF is a constant%
\footnote{For reasons of readability, we have omitted the prefix \code{SPI\_IOC\_} for all \ac{spi} constants from the \spidev \acs{hal}-\acs{api} and we have abbreviated the \ac{spi} constant \code{MESSAGE} with \code{MSG} (also see \autoref{tab:hal:spidev-api}, \ref{tab:specification:spidev-dependencies} and \ref{tab:specification:spidev-thads}).}
that is declared in the \acs{hal}-\acs{api} and \code{LSB\_FIRST} is a pointer to the boolean value \emph{true}.
Using this constant and value always has the effect of setting the bit order to `left-most bit first', independent from the make of the \ac{spi} controller which may follow the de-facto standard or not; this knowledge is part of the device driver and hidden from the application programmer.

\begin{table}[b]
  \caption{Routines of the the \spidev \acs{hal}-\acs{api}.}
  \label{tab:hal:spidev-api}
  \centering
  \begin{tabularx}{\linewidth}{l@{}X}
    \toprule
    \belowrulesepcolor{bgray}
    \rowcolor{bgray}\textbf{\acs{hal}-\acs{api} routine} & \textbf{Provided functionality} \\
    \aboverulesepcolor{bgray}
    \midrule
    \halApiRoutineOpenDR                                 & Initialize peripheral \\
    \halApiRoutineReadFD                                 & Half-duplex read \\
    \halApiRoutineWriteFD                                & Half-duplex write \\
    \halApiRoutineCloseF                                 & Dismiss peripheral \\
    \midrule
    \halApiRoutineIoctlMsgFD                             & Full-duplex transfer \\
    \midrule
    \halApiRoutineIoctlRWModeFD                          & Get/set \acs{cpol}, \acs{cpha} \\
    \halApiRoutineIoctlRWModeExtFD                       & Same as previous but with \qty{32}{\bit} encoded mode to get/set more \acs{spi} config.\@ parameters \\
    \halApiRoutineIoctlRWLSBFFD                          & Get/set \acs{lsbfe} \\
    \halApiRoutineIoctlRWBitsPWordFD                     & Get/set Word~Size \\
    \halApiRoutineIoctlRWSpeedFD                         & Get/set \acs{sclk}~freq. \\
    \bottomrule
  \end{tabularx}
\end{table}

\subsection{Temporal dependencies in the \spidev \acs{hal}-\acs{api}}
\label{sec:spi:thads}
A program can call the routines of the \ac{spi} \ac{hal}-\ac{api} in the Linux kernel (as described in the previous section) in any order.
But only some orders guarantee successful data transmission (assuming the absence of hardware faults).
These orders can be described by dependencies between routines:
If a routine should successfully complete its tasks (in absence of hardware faults), then one or more other routines need to have been successfully completed beforehand.
An obvious example is data transfer with routines \halApiRoutineReadS, \halApiRoutineWriteS, or \halApiRoutineIoctlS:
If these routines are called with an invalid file descriptor, then they will not successfully complete the data transfer.
To have a chance for successful data transfer, routine \halApiRoutineOpenS needs to have successfully completed its task before the call to the data transfer routines.
Note that we refer to \emph{a chance} above because successful data transfer can in general not be guaranteed:
The involved \ac{spi} hardware may fail due an over-voltage or the whole operating system may be in an invalid state due to programming errors that are completely unrelated to the \ac{spi} layers.
Without successfully initializing the \ac{spi} controller and setting up a file descriptor before the data transfer, there is strictly speaking not even a chance for successful data transfer.

The case that, \eg, routine \halApiRoutineReadS \emph{depends} on \halApiRoutineOpenS is represented in the middle block of \autoref{tab:specification:spidev-dependencies}.
Routine \halApiRoutineOpenS in contrast does not depend on other routines from the \ac{spi} \ac{api} because in this paper we disregard for simplicity the case of opening a file, properly closing it, and opening it again.

\begin{table}[b]
  \centering
  \caption{Dependencies in the \spidev \acs{hal}-\acs{api}.}
  \label{tab:specification:spidev-dependencies}
  \begin{tabularx}{\linewidth}{Xll}
    \toprule
    \belowrulesepcolor{bgray}
    \rowcolor{bgray} \textbf{\acs{hal}-\acs{api} routine} & \multicolumn{2}{l}{\textbf{Depends on}} \\
    \aboverulesepcolor{bgray}
    \midrule
    \halApiRoutineOpenDR                      & \multicolumn{2}{l}{---} \\
    \midrule
                                              & \multicolumn{2}{l}{\halApiRoutineOpenD} \\
    \halApiRoutineReadFD                      & \multicolumn{2}{l}{\halApiRoutineIoctlWModeExtFD} \\
    \halApiRoutineWriteFD                     & \multicolumn{2}{l}{\halApiRoutineIoctlWLSBFFD} \\
    \halApiRoutineIoctlMsgFD                  & \multicolumn{2}{l}{\halApiRoutineIoctlWBitsPWordFD} \\
                                              & \multicolumn{2}{l}{\halApiRoutineIoctlWSpeedFD} \\
    \midrule
    \multicolumn{2}{l}{\halApiRoutineCloseF}             & \multirow{6}{*}{\halApiRoutineOpenD} \\
    \multicolumn{2}{l}{\halApiRoutineIoctlRWModeFD}      & \\
    \multicolumn{2}{l}{\halApiRoutineIoctlRWModeExtFD}   & \\
    \multicolumn{2}{l}{\halApiRoutineIoctlRWLSBFFD}      & \\
    \multicolumn{2}{l}{\halApiRoutineIoctlRWBitsPWordFD} & \\
    \multicolumn{2}{l}{\halApiRoutineIoctlRWSpeedFD}     & \\
    \bottomrule
  \end{tabularx}
\end{table}

Interestingly, routine \halApiRoutineReadS also depends on a set of \halApiRoutineIoctlS routines.
The reason is that, next to the issues with standardization as discussed in \autoref{sec:spi:hw}, there is also no universally agreement on default configuration for \ac{spi} controllers.
For example, the default \ac{sclk} frequency value of the local \ac{spi} peripheral integrated into the hardware of a \raspberryPi single-board computer is set to \qty{125}{\mega\hertz}, but many dedicated \ac{spi} peripherals only support a lower \ac{sclk} frequency, \eg, the ADXL345 \ac{spi} slave accelerometer from Analog~Devices supports a maximum \ac{sclk} frequency of \qty{5}{\mega\hertz}.
Hence any attempt to read data without prior configuration of the \ac{sclk} frequency is bound to fail.

Consequently, all routines on the left-hand side of the middle block of \autoref{tab:specification:spidev-dependencies} depend on \halApiRoutineOpenS as well as on four \halApiRoutineIoctlS routines which set the parameters of the \ac{spi} controller.%
\footnote{In this work, the \acs{hal}-\acs{api} routine \halApiRoutineIoctlWModeExt is chosen as dependent routine because it subsumes the functionality of the legacy routine \halApiRoutineIoctlWMode.}
Setting (or getting) \ac{spi} parameters through \halApiRoutineIoctlS of course depends on a previous call of \halApiRoutineOpenS.
These dependencies are captured in the bottom block of \autoref{tab:specification:spidev-dependencies}.

Even worse, misconfigurations of \ac{spi} controllers may in general not only inhibit data transfer but may damage or destroy the sensitive hardware.
One operation mode (next to the star topology as shown in \autoref{fig:spi:star-topology}) is the so-called \num{3}-wire mode where the pin of the omitted \ac{mosi} connection is reused as an input.
If the \ac{spi} controller is not configured for \num{3}-wire mode, then this input pin does not necessarily retain a high impedance and the circuit may be damaged or destroyed if the connected pin is configured as an output (and if there is no explicit protection in place).
Such fatal errors can be avoided if all three data transfer routines depend on the \acs{hal}-\acs{api} routine \halApiRoutineIoctlWModeExt which forces the explicit setting of a \codeParam{mode} parameter value.
This dependency can be refined to not only enforce a performed \halApiRoutineIoctlWModeExt \acs{hal}-\acs{api} routine call but also to involve the \codeParam{mode} parameter and set them to the \ac{spi} topology-specific \codeParam{SPI\_3WIRE} option.
This refinement of the dependency is not modeled in \autoref{tab:specification:spidev-dependencies}, but avoids setting an invalid \codeParam{mode} parameter value, even if the \acs{hal}-\acs{api} routine \halApiRoutineIoctlWModeExt was previously called.
The same applies to the \codeParam{SPI\_NO\_CS} option if the local \ac{spi} peripheral supports the reuse of unused \ac{ss} pins.

Note that the dependencies on \halApiRoutineIoctlS that we have reported here only become known to the embedded software developer after careful study of the \acs{hal}-\acs{api} documentation, the data-sheets of the involved hardware, and the particular hardware setup the program is designed for.
There is no comprehensive internal protection in place against \acs{hal}-\acs{api} misuses that would warn the programmer at compile- or run-time.
There is some protection in the \ac{posix} layer (\eg, all routines that use file descriptors have a defined behavior when called with an invalid file descriptor) but not for the intricacies of the \ac{spi} bus.
Even though knowledge of this kind can be expected of developers, today's informal description of temporal dependencies between \acs{hal}-\acs{api} routines puts an unnecessary burden on them.
When working on the level of `Program' in \autoref{fig:spi:hal-spidev-layers}, they are primarily embedded software developers responsible for developing the application program.

\section{Checking \aclp{thad}}
\label{sec:thad-checking}

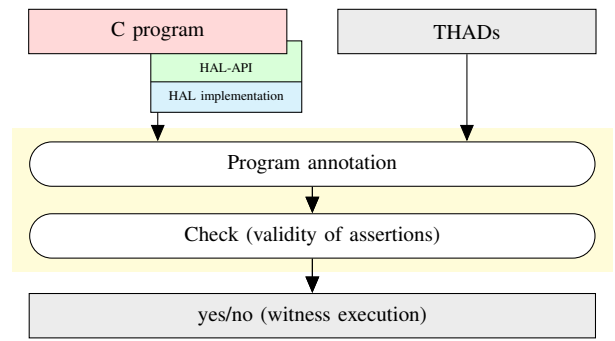
\begin{figure}[b]
  \centering
  \begin{tikzpicture}[
        node distance = 10pt,
        input/.style = {draw, rectangle, minimum width=0.3875\linewidth, minimum height=0.025\textheight, fill=cgray!15},
        output/.style = {input, minimum width=0.85\linewidth, fill=cgray!15},
        inputSub/.style = {minimum width=0.225\linewidth, minimum height=0.01\textheight, anchor=north west, font=\tiny, text opacity=1},
        inputSubFill/.style = {fill opacity=\bFillOpacity},
        task/.style = {draw, fill=white, rounded rectangle, minimum width=0.875\linewidth, minimum height=0.025\textheight},
        pics/inpBoxProg/.style args = {#1/#2/#3}{
            code = {
                \node[input, fill=cred!15] (-box) {#1};
                \begin{pgfonlayer}{behind}
                    \node[inputSub, xshift=-0.225\linewidth+0.0075\textheight] (-halApi) at (-box.south east) {#2};
                    \node[inputSub] (-halImpl) at (-halApi.south west) {#3};
                    \draw ($(-halApi.north west)+(0,0.0075\textheight)$) -- ($(-halApi.north east)+(0,0.0075\textheight)$) -- (-halImpl.south east) -- (-halImpl.south west) -- cycle;
                    \draw (-halApi.south west) -- (-halApi.south east);
                \end{pgfonlayer}
                \begin{pgfonlayer}{background}
                    \fill[cgreen, inputSubFill] ($(-halApi.north west)+(0,0.0075\textheight)$) -- ($(-halApi.north east)+(0,0.0075\textheight)$) -- (-halApi.south east) -- (-halApi.south west);
                    \fill[cblue, inputSubFill] (-halApi.south west) -- (-halApi.south east) -- (-halImpl.south east) -- (-halImpl.south west);
                \end{pgfonlayer}
            }
        },
        pics/inpBoxProg/.default=C~program/\acs{hal}-\acs{api}/\acs{hal}~implementation,
    ]
    \pic (inpProg) {inpBoxProg};
    \node[input, right = 0.075\linewidth of inpProg-box] (inpThads) {\acsp{thad}};
    \coordinate (inpMiddle) at ($(inpProg-box.south)!0.5!(inpThads.south)$);
    \node[task, below = 33pt of inpMiddle] (annotProg) {Program annotation};
    \node[task, below = of annotProg] (verifProg) {Check (validity of assertions)};
    \node[output, below = 13pt of verifProg] (verifResult) {yes/no (witness execution)};
    \coordinate (chelp) at (inpProg-box |- inpProg-halImpl.south);
    \draw[arrow] (chelp) -- (chelp |- annotProg.north);
    \draw[arrow] (inpThads) -- (inpThads |- annotProg.north);
    \draw[arrow] (annotProg) -- (verifProg);
    \draw[arrow] (verifProg) -- (verifResult);
    \draw[arrow] (annotProg) -- (verifProg);
    \draw[arrow] (verifProg) -- (verifResult);
    \coordinate (fillThadVerifNorthWest) at ($(annotProg.north west) + (-0.015\textheight-5pt, 5pt)$);
    \coordinate (fillThadVerifSouthEast) at ($(verifProg.south east) + (0.015\textheight+5pt,-5pt)$);
    \begin{pgfonlayer}{background}
        \fill[cyellow, opacity = \bFillOpacity] (fillThadVerifNorthWest) rectangle (fillThadVerifSouthEast);
    \end{pgfonlayer}
\end{tikzpicture}
  \caption{Overall approach for checking the validity of given \acsp{thad} in a given C~program that calls \acs{hal}-\acs{api} routines.}
  \label{fig:thads:verification-process}
\end{figure}

Given a C~program and a set of dependencies for the \acs{hal}-\acs{api}, the developer will check whether each execution of the C~program obeys to the dependencies (\ie, the check will detect possible violations of dependencies or, respectively, give a formal guarantee of their absence); see \autoref{fig:thads:verification-process}.

This section is structured as follows.
We first convey an intuition for the concepts of a \ac{thad} and a program annotation by means of example in the context of the \spidev \acs{hal}-\acs{api}.
We then discuss \acp{thad} and program annotations in general and we explain the general principle behind the \emph{software model checking} algorithm for implementing the check.

\subsection{Example of the \spidev \acs{hal}-\acs{api}}

\begin{table}
  \caption{Set of \acp{thad} formalizing \spidev \acs{hal}-\acs{api} dependencies from \autoref{tab:specification:spidev-dependencies}.
           The dependency $\delta_1$ (``\halApiRoutineReadS depends on \halApiRoutineOpenS'') expresses that a call of the routine \halApiRoutineReadS must not occur without a preceding call of the routine \halApiRoutineOpenS.}
  \label{tab:specification:spidev-thads}
  \centering
  \setlength{\tabcolsep}{1pt}
  \newcommand{\varThadDepRow}[3]{ #1 & $:$ & #2 & $\opdep$ & #3 \\ }
  \begin{tabularx}{\linewidth}{lcXcX}
    \toprule
    \varThadDepRow{$\varThad_{1}$}{$\halApiRoutineOpenS$}{$\halApiRoutineReadS$}
    \varThadDepRow{$\varThad_{2}$}{$\halApiRoutineOpenS$}{$\halApiRoutineWriteS$}
    \varThadDepRow{$\varThad_{3}$}{$\halApiRoutineOpenS$}{$\halApiRoutineIoctlMsg$}
    \varThadDepRow{$\varThad_{4}$}{$\halApiRoutineOpenS$}{$\halApiRoutineCloseS$}
    \varThadDepRow{$\varThad_{5}$}{$\halApiRoutineOpenS$}{$\halApiRoutineIoctlRMode$}
    \varThadDepRow{$\varThad_{6}$}{$\halApiRoutineOpenS$}{$\halApiRoutineIoctlWMode$}
    \varThadDepRow{$\varThad_{7}$}{$\halApiRoutineOpenS$}{$\halApiRoutineIoctlRModeExt$}
    \varThadDepRow{$\varThad_{8}$}{$\halApiRoutineOpenS$}{$\halApiRoutineIoctlWModeExt$}
    \varThadDepRow{$\varThad_{9}$}{$\halApiRoutineOpenS$}{$\halApiRoutineIoctlRLSBF$}
    \varThadDepRow{$\varThad_{10}$}{$\halApiRoutineOpenS$}{$\halApiRoutineIoctlWLSBF$}
    \varThadDepRow{$\varThad_{11}$}{$\halApiRoutineOpenS$}{$\halApiRoutineIoctlRBitsPWord$}
    \varThadDepRow{$\varThad_{12}$}{$\halApiRoutineOpenS$}{$\halApiRoutineIoctlWBitsPWord$}
    \varThadDepRow{$\varThad_{13}$}{$\halApiRoutineOpenS$}{$\halApiRoutineIoctlRSpeed$}
    \varThadDepRow{$\varThad_{14}$}{$\halApiRoutineOpenS$}{$\halApiRoutineIoctlWSpeed$}
    \midrule
    \varThadDepRow{$\varThad_{15}$}{$\halApiRoutineIoctlWModeExt$}{$\halApiRoutineReadS$}
    \varThadDepRow{$\varThad_{16}$}{$\halApiRoutineIoctlWModeExt$}{$\halApiRoutineWriteS$}
    \varThadDepRow{$\varThad_{17}$}{$\halApiRoutineIoctlWModeExt$}{$\halApiRoutineIoctlMsg$}
    \varThadDepRow{$\varThad_{18}$}{$\halApiRoutineIoctlWLSBF$}{$\halApiRoutineReadS$}
    \varThadDepRow{$\varThad_{19}$}{$\halApiRoutineIoctlWLSBF$}{$\halApiRoutineWriteS$}
    \varThadDepRow{$\varThad_{20}$}{$\halApiRoutineIoctlWLSBF$}{$\halApiRoutineIoctlMsg$}
    \varThadDepRow{$\varThad_{21}$}{$\halApiRoutineIoctlWBitsPWord$}{$\halApiRoutineReadS$}
    \varThadDepRow{$\varThad_{22}$}{$\halApiRoutineIoctlWBitsPWord$}{$\halApiRoutineWriteS$}
    \varThadDepRow{$\varThad_{23}$}{$\halApiRoutineIoctlWBitsPWord$}{$\halApiRoutineIoctlMsg$}
    \varThadDepRow{$\varThad_{24}$}{$\halApiRoutineIoctlWSpeed$}{$\halApiRoutineReadS$}
    \varThadDepRow{$\varThad_{25}$}{$\halApiRoutineIoctlWSpeed$}{$\halApiRoutineWriteS$}
    \varThadDepRow{$\varThad_{26}$}{$\halApiRoutineIoctlWSpeed$}{$\halApiRoutineIoctlMsg$}
    \bottomrule
  \end{tabularx}
\end{table}

To formalize the dependencies in the \spidev \acs{hal}-\acs{api} from \autoref{tab:specification:spidev-dependencies}, the developer can formulate the \acp{thad} in \autoref{tab:specification:spidev-thads}, and generate the annotation of the C~program automatically.

Alternatively, the developer can write the annotation manually; see \autoref{lst:annotation:c-thad-single} and~\ref{lst:annotation:c-thad-multiple} for a glimpse of the annotation corresponding to the dependencies in the \spidev \acs{hal}-\acs{api} from \autoref{tab:specification:spidev-dependencies} or, equivalently, to the \acp{thad} in \autoref{tab:specification:spidev-thads}.
The annotation consists of \emph{auxiliary program statements} inserted in the code for the \ac{hal} implementation of \spidev.
The code snippets here refer to the original \ac{hal} implementation of the \acs{hal}-\acs{api} routines \halApiRoutineOpenS, \halApiRoutineIoctlS, and \halApiRoutineReadS.

\begin{codelisting}
  \setlength{\fboxsep}{1pt}
\begin{lstlisting}[mathescape]
<@\colorbox{green!20}{/*@\ \textbf{ghost} \textbf{int} state\_d3 = 0; */}@>

int open(const char *path, int oflag, ...) {
    int ret = $\dots$;

    <@\colorbox{green!20}{/*@\ \textbf{ghost} state\_d3 = 1; */}@>
    return ret;
}

int ioctl(int fd, int request, ...) {
    if (request == MSG) {
        <@\colorbox{green!20}{/*@\ \textbf{assert} (state\_d3 == 1); */}@>
    }

    return $\dots$;
}
\end{lstlisting}
  \caption{Annotation (in green) corresponding to one \acs{thad}, here the \acs{thad}~$\varThadDep{\varThad_{3}}{\halApiRoutineOpenS}{\halApiRoutineIoctlMsg}$ from \autoref{tab:specification:spidev-thads}.}
  \label{lst:annotation:c-thad-single}
\end{codelisting}

\begin{codelisting}
  \setlength{\fboxsep}{1pt}
\begin{lstlisting}[mathescape]
<@\colorbox{green!20}{/*@\ \textbf{ghost} \textbf{int} state\_d1\ \ = 0; */}@>
<@\colorbox{green!20}{/*@\ \textbf{ghost} \textbf{int} state\_d8\ \ = 0; */}@>
<@\colorbox{green!20}{/*@\ \textbf{ghost} \textbf{int} state\_d15\ = 0; */}@>

int open(const char *path, int oflag, ...) {
    int ret = $\dots$;

    <@\colorbox{green!20}{/*@\ \textbf{ghost} state\_d1 = 1; */}@>
    <@\colorbox{green!20}{/*@\ \textbf{ghost} state\_d8 = 1; */}@>
    return ret;
}

int ioctl(int fd, int request, ...) {
    if (request == WR_MODE32) {
        <@\colorbox{green!20}{/*@\ \textbf{assert}(state\_d8 == 1); */}@>
    }

    int ret = $\dots$;

    <@\colorbox{green!20}{/*@\ \textbf{ghost} state\_d15 = 1; */}@>
    return ret;
}

ssize_t read(int fd, void *buf, size_t nbyte) {
    <@\colorbox{green!20}{/*@\ \textbf{assert}(state\_d1\ \ == 1); */}@>
    <@\colorbox{green!20}{/*@\ \textbf{assert}(state\_d15\ == 1); */}@>

    return $\dots$;
}
\end{lstlisting}
  \caption{Annotation (in green) corresponding to a set of \acsp{thad}, here the \acsp{thad}~$\varThad_{1}$, $\varThad_{8}$, and~$\varThad_{15}$ from \autoref{tab:specification:spidev-thads}.}
  \label{lst:annotation:c-thad-multiple}
\end{codelisting}

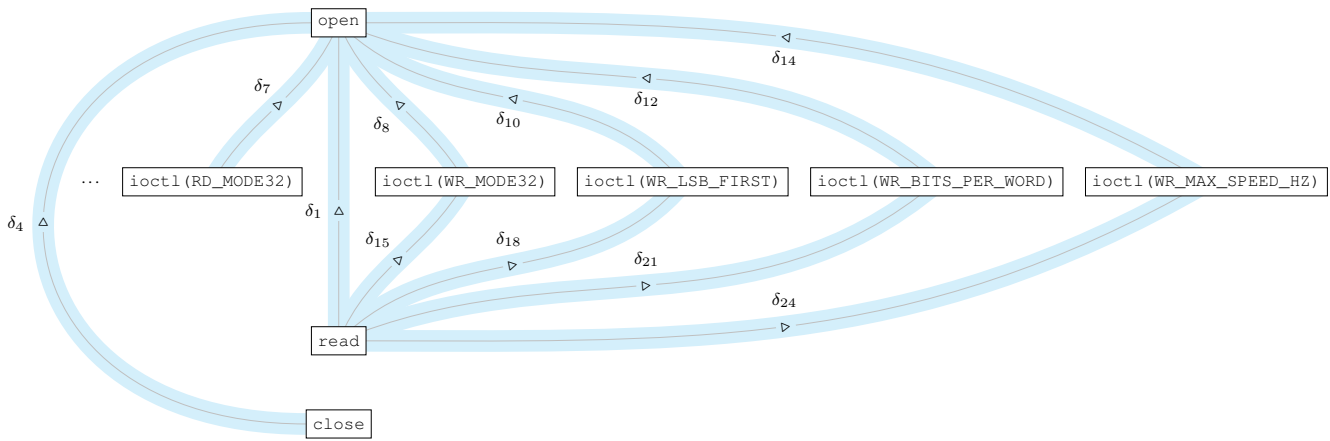
\begin{figure*}
  \centering
  \resizebox{\textwidth}{!}{
    \begin{tikzpicture}[node distance = 60pt and 10pt]
    \clip (-5.5cm,-7cm) rectangle (16.25cm,0.5cm);
    \node[thadRoutineBox] (open) {$\halApiRoutineOpenS$};
    \node[below=of open,thadRoutine] (cmiddle) {};
    \node[right=of cmiddle,thadRoutineBox] (wmode) {$\halApiRoutineIoctlWModeExt$};
    \node[right=of wmode,thadRoutineBox] (wlsbf) {$\halApiRoutineIoctlWLSBF$};
    \node[right=of wlsbf,thadRoutineBox] (wbpw) {$\halApiRoutineIoctlWBitsPWord$};
    \node[right=of wbpw,thadRoutineBox] (wspeed) {$\halApiRoutineIoctlWSpeed$};
    \node[below=of cmiddle,thadRoutineBox] (read) {$\halApiRoutineReadS$};
    \node[left=of cmiddle,thadRoutineBox] (rmode) {$\halApiRoutineIoctlRModeExt$};
    \node[left=0.2 of rmode] (r) {\dots};
    \node[below=25pt of read,thadRoutineBox] (close) {$\halApiRoutineCloseS$};
    \begin{pgfonlayer}{behind}
        \draw[thad=0.6] (open.south) to node[pos=0.6,left,xshift=-4pt]{$\varThad_{1}$} (read.north);
        \draw[thad] (open.center) to [out=-112.5,in=60] node[above left,yshift=1.5pt]{$\varThad_{7}$} (rmode.center);
        \draw[thad] (open.center) to [out=-67.5,in=120] node[below left,yshift=-2.5pt]{$\varThad_{8}$} (wmode.center);
        \draw[thad] (open.center) to [out=-45,in=130] node[below,xshift=-2.5pt,yshift=-2.5pt]{$\varThad_{10}$} (wlsbf.center);
        \draw[thad] (open.center) to [out=-22.5,in=140] node[below,xshift=-2.5pt,yshift=-2.5pt]{$\varThad_{12}$} (wbpw.center);
        \draw[thad] (open.center) to [out=0,in=150] node[below,xshift=-2.5pt,yshift=-2.5pt]{$\varThad_{14}$} (wspeed.center);
        \draw[thad] (wmode.center) to [out=-120,in=67.5] node[above left,yshift=2.5pt]{$\varThad_{15}$} (read.center);
        \draw[thad] (wlsbf.center) to [out=-130,in=45] node[above,xshift=-2.5pt,yshift=5pt]{$\varThad_{18}$} (read.center);
        \draw[thad] (wbpw.center) to [out=-140,in=22.5] node[above,xshift=-2.5pt,yshift=5pt]{$\varThad_{21}$} (read.center);
        \draw[thad] (wspeed.center) to [out=-150,in=0] node[above,xshift=-2.5pt,yshift=5pt]{$\varThad_{24}$} (read.center);
        \draw[thad] (open.west) to [out=180,in=180,looseness=2.25] node[left,xshift=-5pt]{$\varThad_{4}$} (close.west);
    \end{pgfonlayer}
\end{tikzpicture}
  }
  \caption{Graph of \acsp{thad} for \spidev \acs{hal}-\acs{api} dependencies.
           For example, \halApiRoutineCloseS depends on \halApiRoutineOpenS only, whereas \halApiRoutineReadS depends on \halApiRoutineOpenS and on four other routines with specific parameters (and on several other routines, not shown here).}
  \label{fig:specification:thad-form-spidev}
\end{figure*}

We use the graph shown in \autoref{fig:specification:thad-form-spidev} to depict the \acp{thad} listed in \autoref{tab:specification:spidev-thads}.
For clarity, we show only a subset of all \acp{thad} and omit the details regarding the consistent use of file descriptors.
An interesting by-product of showing the formalization of the \spidev dependencies is to exhibit their (perhaps surprising) complexity.
The complexity stems from the fact there is an exponential number (exponential in the number of \acp{thad}) of interleaving orderings between the different routines that a given routine depends on.

For the developer, this complexity adds to the general difficulty of mastering the complexity of the control flow of the program (where, for example, \halApiRoutineReadS may depend on routines that are called in a loop or in different branches).
The bug in the firmware program of the \trezor hardware wallet is a witness to this difficulty (see \autoref{sec:introduction:examples}).

\subsection{Program annotation}
\label{chp:specification:annotation}
In a formal method, the developer must specify the correctness of the C~program formally, either explicitly or implicitly.
Next, we discuss the two conceptual options for the developer.

One option allows the developer to formalize the \acs{hal}-\acs{api} dependencies by formulating the corresponding \acp{thad}.
Then, given a set of \acp{thad} and given a C~program, the annotation can be generated automatically.
Generating the program annotation reduces the problem of correctness \wrt to an explicitly given specification of correctness to the problem of correctness \wrt implicitly given specification of correctness.
The step of formulating \acp{thad} is independent of any application and can be done even in absence of an implementation of the \ac{hal}.

Another option enables the developer to formalize the correctness (of the given C~program \wrt the \acs{hal}-\acs{api} dependencies) by manually writing the annotations directly.
As we will see, the annotation can be restricted to the part of the code for the \ac{hal} implementation of the used \acs{hal}-\acs{api}; \ie, the annotation thus does not need to touch the application-specific part of the C~program.
In this sense, the formalization of correctness by directly annotating the C~program is also independent of the particular application.
However, this is only possible once the implementation of the \ac{hal} is available.

The manual effort for the annotation does seem manageable: for each \ac{thad}, the developer has to insert two auxiliary program statements (in addition to the declaration of an auxiliary variable).
The annotated C~program contains auxiliary program statements of two kinds: statements that update auxiliary variables, or assert statements, \ie, statements of the form \code{\textbf{assert}(exp)} where \code{exp} stands for a Boolean expression (in program variables, original or auxiliary).
Intuitively, we may translate the statement \code{\textbf{assert}(exp)} to the statement \mbox{\code{\textbf{if} (exp) \textbf{then} skip \textbf{else} fail}}.
The execution of the assert statement \emph{succeeds} if the Boolean expression evaluates to \emph{true}, and it \emph{fails} if the Boolean expression evaluates to \emph{false}.
The annotated program is called \emph{safe} if its assert statements can never fail (\ie, there exists no execution of the C~program where the Boolean expression \code{exp} evaluates to \emph{false} when the corresponding program point is reached).
The correctness of an annotation \wrt a given set of \acp{thad} means that the C~program is correct \wrt the \acp{thad} if and only if the annotated C~program is safe.
We can build up on existing annotation languages, such as the \ac{acsl}, which are widely used and supported by a variety of verification tools.
Note that the annotations use the existing syntax and semantics of program statements.

Independently of how \acp{thad} are employed (whether they are used explicitly or implicitly), \acp{thad} may form a class of correctness properties worthy of independent investigation.
We have formally defined their semantics (to know when an execution of the given C~program obeys the dependencies), and we have proved the correctness of the generated program annotation.
The proof and the formal setup is omitted, as both is not immediately necessary to be presented in this work.

\subsection{Software model checking}
To check the C~program automatically, we use software model checking.
Explained on a high level, the software model checking algorithm consists of the iteration of the \emph{\ac{cegar} loop}.
In each iteration of the loop, the algorithm checks the correctness on an abstraction of the C~program.
If the abstraction is too coarse, the algorithm refines the abstraction for the next iteration.
The abstraction is too coarse if there is a \emph{spurious} counter-example, \ie, an execution which is feasible in the abstraction but infeasible in the concrete C~program.
Analyzing the infeasibility of the counter-example reveals the missing precision (\ie, critical information that got lost in the abstraction and that apparently must be recovered).
The algorithm refines the abstraction to remedy the detected lack of precision.
The refined abstraction will preserve the critical information.
In the next iteration, when the algorithm checks the correctness of the \emph{refined} abstraction, the spurious counter-example is infeasible not only in the concrete C~program but also in the abstraction (\ie, it will no longer appear).

Thus, there are three possible options for the outcome of a check applied to a C~program: the check terminates and returns \emph{yes} (the program is proven correct), the check runs into a timeout (the program may be correct or incorrect), or the check terminates and returns \emph{no}, together with a (feasible) counter-example (the program is incorrect).

\subsection{\acsp{thad} in the embedded software development process}
Once the developer has formalized a \acs{hal}-\acs{api} dependency as a \ac{thad} and the \ac{hal} implementation (or a wrapper) has been annotated by following the procedure outlined in \autoref{chp:specification:annotation}, any embedded program that uses the \acs{hal}-\acs{api} can be checked within the build procedure.
The correctness of the annotations implies that a program that \emph{violates} a \ac{thad} will never be proven correct.
That is, if, for example, the \trezor developers would formalize only the one \acs{usb}-related dependency, they could ensure that they never again deploy binaries that violate the dependency.
In other words, embedded developers can use \acp{thad} to effectively protect themselves against repeating certain errors related to the \acs{hal}-\acs{api}:
each analyzed and fixed bug would increase the set of \acp{thad}.

A complementary approach would work top-down:
An individual embedded developer (or a group leader, or even the company providing the \ac{hal}) would analyze the documentation for dependencies (as we did in \autoref{sec:spi}) and would formally specify a set of \acp{thad} for the \acs{hal}-\acs{api} \emph{before} the first program line is written.
Thereby, \ac{thad} violations would not get past the software model checker and hence the tedious effort of compiling, uploading, testing, and debugging on the target platform would be avoided.

\section{Case Study}
\label{sec:casestudy}

\newcommand{\Pmcp}{I/O Expander}
\newcommand{\Padxl}{Accelerometer}
\newcommand{\Plx}{\spidev-Test}

We evaluate the formalization and checking of \acp{thad} using three programs that utilize the \spidev \ac{hal} as discussed in \autoref{sec:spi}.
These programs vary in size, complexity, and number of relevant \acp{thad}, each representing a different use case of embedded systems programming at the application level (\cf `Program' in \autoref{fig:spi:hal-spidev-layers}).
We apply the \ac{thad} verification method from \autoref{sec:thad-checking} to the three C~programs to assess practical feasibility and resource consumption.

\subsection{Example programs}
\label{sec:evaluation:programs}

\subsubsection*{\Pmcp}
\label{sec:evaluation:program-mcp}
The first program is `\Pmcp'.
It uses the \spidev \acs{hal}-\acs{api} directly to operate a so-called input/output expander via a \raspberryPi single-board computer.
This program has been written for a setup where the Raspberry~Pi connects to a Microchip MCP23S17.
The MCP23S17 provides \num{16}~configurable input/output pins and serves as a representative for a low-level sensor or actuator with an integrated \ac{spi} slave controller.
In this setup, the pins are configured as outputs to switch lamps on and off.

\begin{figure}
  \centering
  \begin{tikzpicture}[
      xscale=5,
      node distance=3pt
    ]
    \pic (app) {swlapp=`\Pmcp{}' and `\Plx'};
    \pic (spidev) [below = of app-app] {swlintend=\spidev~\acs{hal}-\acs{api}/\spidev~\acs{hal}~implementation};
    \pic (hw) [below = of spidev-endhal] {swlhw=\raspberryPi~hardware};
  \end{tikzpicture}
  \caption{\acs{hal} used by `\Pmcp' and `\Plx'.}
  \label{fig:prog13hals}
\end{figure}
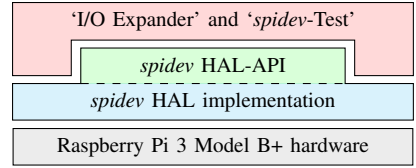

From the hardware topology perspective (\cf \autoref{fig:spi:star-topology}), this setup has one \ac{spi} master controller on the single-board computer and one \ac{spi} slave controller in the MCP23S17.
From the software architecture perspective (\cf \autoref{fig:prog13hals}), program `\Pmcp' uses the \spidev interface directly, which (omitting the layers in-between, \cf \autoref{fig:spi:hal-spidev-layers}) operates the \ac{spi} master controller.
The program needs to open the \ac{spi} device file, to call \halApiRoutineIoctlS in order to configure the \ac{spi} master to match the parameters of the MCP23S17, and to initiate a data transfer from the local \ac{spi} master to the remote \ac{spi} slave.
The transfer includes a control command for the MCP23S17 to turn on the lamp, which is wired to pin~A0.

The program is of moderate size (76~lines including comments, \cf \autoref{tab:evaluation:results}) but not trivial.
All but \acp{thad}~$\varThad_{3}$, $\varThad_{4}$, $\varThad_{14}$, and~$\varThad_{26}$ from \autoref{tab:specification:spidev-thads} are trivially satisfied, as the program does not call any depending routine of those \acp{thad}.

\subsubsection*{\Padxl}
\label{sec:evaluation:program-adxl}
The second program is `\Padxl'.
It is supposed to transfer control commands from a local \ac{spi} master to a remote \ac{spi} slave and in order to receive data from the remote \ac{spi} slave.
The remote \ac{spi} slave is the accelerometer ADXL345 from Analog Devices Inc.\@ which measures the acceleration on three axis.
The ADXL345 accelerometer is a representative for a versatile sensor with an integrated \ac{spi} slave controller.
This program is intended to be executed on the Raspberry~Pi model mentioned above.

\begin{figure}
  \centering
  \begin{tikzpicture}[
      node distance=3pt
    ]
    \pic (app) {swlapp=`\Padxl'};
    \pic (adxlhal) [below = of app-app] {swlintst=ADXL345~\acs{hal}-\acs{api}/ADXL345~\acs{hal}~implementation};
    \pic (spidev) [below = of adxlhal-midhal] {swlintend=\spidev~\acs{hal}-\acs{api}/\spidev~\acs{hal}~implementation};
    \pic (hw) [below = of spidev-endhal] {swlhw=\raspberryPi~hardware};
  \end{tikzpicture}
  \caption{\acsp{hal} used by C~program `\Padxl'.}
  \label{fig:prog2hals}
\end{figure}
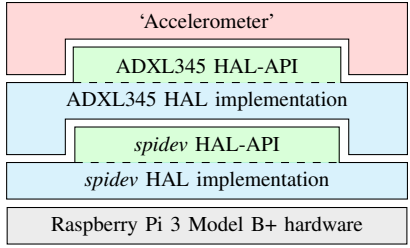

The C~program does not use the \spidev \acs{hal}-\acs{api} directly as shown in \autoref{fig:prog2hals}, but instead uses the \acs{api} of the ADXL345 library provided by Analog Devices Inc.%
\footnote{\acs{api} and \acs{hal} implementation of the ADXL345 library: \url{https://github.com/analogdevicesinc/no-OS/tree/master/drivers/accel/adxl345}.
          Note that the C~program `\Padxl' uses the ADXL345 library in version of the \href{https://github.com/analogdevicesinc/no-OS/tree/30f903bfa74cc24d3fe1b97a07d93682e45daf2b/drivers/accel/adxl345}{commit 30f903bfa74cc24d3fe1b97a07d93682e45daf2b}.}
The \acs{api} of the ADXL345 \ac{hal} provides functionality for the ADXL345 accelerometer to generate valid control commands and to process received raw data.
The ADXL345 \ac{hal} implementation uses the \spidev \acs{hal}-\acs{api} to transfer control commands from the local \ac{spi} master to the remote ADXL345 \ac{spi} slave and to receive requested acceleration data.
Hence this program is a representative for the case where a \acs{hal}-\acs{api} is used \emph{indirectly} but where the programmer wants to ensure that all temporal dependencies of this \ac{hal} are respected.

The program `\Padxl' is of substantially larger size than the first program (316~lines including comments, \cf \autoref{tab:evaluation:results}), partly due to the case that the ADXL345 library has been inlined for checking purposes.
The program needs to satisfy the same \acp{thad} as the previous program, and additionally the \acp{thad} $\varThad_{8}$ and~$\varThad_{17}$ from \autoref{tab:specification:spidev-thads}.
Note that the \acp{thad}~$\varThad_{8}$ and $\varThad_{17}$ appear in parentheses in \autoref{tab:evaluation:results} since the `\Padxl' program calls the legacy \acs{hal}-\acs{api} routine \halApiRoutineIoctlWMode instead of the routine \halApiRoutineIoctlWModeExt, which is involved in~$\varThad_{8}$ and~$\varThad_{17}$.
Since \halApiRoutineIoctlWMode offers the same functionality as \halApiRoutineIoctlWModeExt, we will replace \halApiRoutineIoctlWModeExt in~$\varThad_{8}$ and~$\varThad_{17}$ with \halApiRoutineIoctlWMode for the annotation of this program.

Note that the program `\Padxl' directly corresponds to the second example that we have discussed in \autoref{sec:introduction:examples} and hence exists in two versions:
A faulty version which contains the error described in \autoref{sec:introduction:examples} (a missing configuration routine call, leasing to sporadic system failures), and one version that satisfies all relevant \acp{thad}.

\subsubsection*{\Plx}
\label{sec:evaluation:program-lx}
The third and last program is `\Plx'.
This program has directly been taken from the source code repository of the Linux~5.3 kernel.%
\footnote{Linux~5.3 `\Plx' program: \url{https://git.kernel.org/pub/scm/linux/kernel/git/torvalds/linux.git/tree/tools/spi/spidev_test.c?h=v5.3}.}
It provides Linux developers with test cases to validate newly developed components, such as \ac{spi} device drivers, against the Linux kernel's \ac{spi} subsystem.
The `\Plx' program allows a test engineer to specify \ac{spi} configuration parameters and to parameterize a data transfer from a local \ac{spi} peripheral to a remote \ac{spi} peripheral.
The program tries to set and get \ac{spi} configuration parameters of a local \ac{spi} peripheral and performs the parameterized data transfer while measuring the data throughput.
This program is a representative for an embedded program that makes extensive use of the \spidev \acs{hal}-\ac{api}.

The program is the largest of the three (481~lines including comments, \cf \autoref{tab:evaluation:results}) and uses the \spidev \acs{hal}-\acs{api} directly, like `\Pmcp' (see \autoref{fig:prog13hals}).
In addition to the \acp{thad} mentioned above, it needs to satisfy the \acp{thad}~$\varThad_{7}$, $\varThad_{11}$ to $\varThad_{13}$, and $\varThad_{23}$ from \autoref{tab:specification:spidev-thads}, the others are trivially satisfied.

\subsection{Treating \acsp{thad} with data dependencies.}
\label{sec:evaluation:thads-data-dependencies}
Recall that the presentation in \autoref{sec:thad-checking} omitted the treatment of the data dependency expressed by the shared file descriptor for brevity and simplicity.
For example, \ac{thad}~$\varThad_{1}$ spelled out actually reads $\varThadDep{\varThad_{1}}{\halApiRoutineOpenDR}{\halApiRoutineReadF}$ (``using \halApiRoutineReadF on file $\mathit{fd}$ needs a previous completion of routine \halApiRoutineOpenD which returned  $\mathit{fd}$'').
This refined level of detail can immediately be supported by introducing a second ghost variable $\code{fd\_d1}$ of type \emph{file descriptor} and ghost code inside \halApiRoutineOpenS which assigns the chosen file descriptor to $\code{fd\_d1}$.
An additional assertion in \halApiRoutineReadF then checks whether the value of parameter $\code{fd}$ matches $\code{fd\_d1}$.
In general, \acp{thad} for the \spidev \ac{hal}-\ac{api} also needs to ensure that a device file of the local \ac{spi} peripheral is opened, and that the access mode is correctly set.
In our experiments, we have used a fixed file path and a fixed mode, which requests read and write access.

\subsection{Results}
\label{sec:evaluation:results}

\begin{table}[b]
  \centering
  \renewcommand{\cmark}{$\bullet$}
  \newcommand{\loc}[2]{#1\,/\,#2~\acs{loc}}
  \caption{Program characteristics (lines of code with/without annotations) and checking results (\acs{thad} checked and satisfied `\protect\cmark', and checking time and memory consumption).}
  \label{tab:evaluation:results}
  \begin{tabularx}{\linewidth}{XlcccX}
    \toprule
    \belowrulesepcolor{bgray}
    \rowcolor{bgray} & & \Pmcp & \Padxl & \Plx & \\
    \aboverulesepcolor{bgray}
    \midrule
    & prog.\@ size & \loc{76}{181} & \loc{316}{594} & \loc{481}{766} & \\
    \midrule
    & $\varThad_{3}$  & \cmark & \cmark & \cmark & \\
    & $\varThad_{4}$  & \cmark & \cmark & \cmark & \\
    & $\varThad_{7}$  &        &        & \cmark & \\
    & $\varThad_{8}$  &        &(\cmark)& \cmark & \\
    & $\varThad_{11}$ &        &        & \cmark & \\
    & $\varThad_{12}$ &        &        & \cmark & \\
    & $\varThad_{13}$ &        &        & \cmark & \\
    & $\varThad_{14}$ & \cmark & \cmark & \cmark & \\
    & $\varThad_{17}$ &        &(\cmark)& \cmark & \\
    & $\varThad_{23}$ &        &        & \cmark & \\
    & $\varThad_{26}$ & \cmark & \cmark & \cmark & \\
    \midrule
    & time   & \qty{4.15}{\second}   & \qty{11.64}{\second}  & \qty{54.94}{\second}  & \\
    & memory & \qty{340}{\mega\byte} & \qty{541}{\mega\byte} & \qty{955}{\mega\byte} & \\
    \bottomrule
  \end{tabularx}
\end{table}

We have applied the program annotation as outlined in \autoref{sec:thad-checking} to all three programs which increases the program sizes to 181, 594, and 766~lines, respectively, including generously added comments.
The software model checker \ultimateAutomizer~\cite{Heizmann2018} in version~0.1.25-a62cfef8 is able to prove the validity of all assertions that have been added by the annotation procedure.
Hence we can conclude that all three programs satisfy all \acp{thad} (of `\Padxl' we only checked the correct one; see below).
\autoref{tab:evaluation:results} indicates with bullet points which \ac{thad} is relevant for which program, all of them were found to be satisfied.
Note that \autoref{tab:evaluation:results} does not contain all observed \acp{thad} from the \spidev \acs{hal}-\acs{api}.
For evaluation purposes, we have for each program only considered those \acp{thad} whose depending routine is called in the program (otherwise, the \ac{thad} is trivially satisfied, see above).
Hereby, we have emulated the consideration of application programs with different numbers of \acp{thad}.

The last two rows of \autoref{tab:evaluation:results} report execution time and memory consumption as measured on the computer that was used for checking the \acp{thad}.%
\footnote{Quad-core CPU at \qty{3.4}{\giga\hertz}, \qty{8}{\giga\byte} memory, Arch~Linux with kernel~5.6.3.}
We see that the verification time with \ultimateAutomizer is well below one minute for the typical embedded programs `\Pmcp' and `\Padxl', and that even the high complexity example `\Plx' can be checked within \qty{55}{\second}.
These checking times seem to be acceptable for an embedded systems workflow.
Detecting the error in the faulty version of `\Padxl' (see \autoref{sec:evaluation:program-adxl}) takes less than one second.
The times reported in \autoref{tab:evaluation:results} do not include the time that is necessary to complete the program annotation, but only the total time of a verification run with \ultimateAutomizer.
Firstly, as explained in \autoref{sec:thad-checking}, the program annotation for \acp{thad} only takes place in the \ac{hal} implementation of a \acs{hal}-\acs{api} and hence needs to be done only once (or seldomly) to then be reused in any C~program which uses that \acs{hal}-\acs{api}.
Secondly, the program annotation is a simple source-to-source transformation (both, the input and the output, of the procedure are C~programs), hence the execution times of implementations of the program annotation can be expected to be neglectable.
The peak memory consumption of all program verifications with \ultimateAutomizer never exceeds \qty{1}{\giga\byte}, where a verification of the annotated C~program `\Plx' required a maximum memory consumption of \qty{955}{\mega\byte}.

Overall, we observe that a verification result is available within a reasonable time on a standard desktop computers commonly used by embedded system developers.
There is no need for dedicated server hardware or alternative formal verification methods.
The program `\Padxl' demonstrates that the checking of \acp{thad} is applicable not only to C~programs that use a \acs{hal}-\acs{api} \emph{directly}, but also to third-party libraries that use such a \acs{hal}-\acs{api}.
This capability is particularly valuable for checking \acp{thad} in third-party libraries after their publication, especially when a software developer bears full liability for a safety-critical system.
In particular, `\Plx' illustrates that the checking of \acp{thad} can assist test engineers in avoiding incorrectly implemented test cases that could cause false negatives.
Test engineers can apply the checking approach to their test programs that use \acs{hal}-\acsp{api} to detect \ac{thad} violations in test case implementations.

\section{Conclusion and Future Work}
\label{sec:conclusion}
We have introduced an approach to formulate \acp{thad} for embedded software (at the application layer) and verify them automatically using software model checking.
Our main contribution of this work is to single out a specific class of correctness properties for a particular class of programs and demonstrate that we thus obtain an interesting application domain where a formal method based on software model checking has a promising potential.

Our case study indicates that the approach is practically feasible on realistic examples.
The next step is to explore scalability.
Specifically, the question here is how we can systematically increase the difficulty for the developer in applying the approach.
The difficulty relates to the complexity of the \acp{thad} itself (which is perhaps more than just the number of \acp{thad}) and to the complexity of the control flow (which is perhaps more than the number of lines of code).

A motivation for focusing on the class of \acp{thad} is its \emph{minimality} by design (just above generic correctness properties like array bounds).
An open question is whether other minimal, yet relevant, classes of correctness properties (possibly in different contexts) could benefit from similar investigations.

Future work could also explore methods, both manual and automated, for eliciting \acp{thad} and integrating them seamlessly into the embedded software development lifecycle.

\section*{Acknowledgments}
Partially funded by the Deutsche Forschungsgemeinschaft (DFG, German Research Foundation) -- 503812980.

\bibliographystyle{IEEEtran}
\bibliography{references.bib}

@Article{Anderson2008,
  author  = {Anderson, Paul},
  journal = {CrossTalk: The Journal of Defense Software Engineering},
  title   = {The Use and Limitations of Static-Analysis Tools to Improve Software Quality},
  year    = {2008},
  number  = {6},
  pages   = {18--21},
  volume  = {21},
}

@InProceedings{AbbaspourAsadollah2015,
  author    = {Abbaspour Asadollah, Sara and Inam, Rafia and Hansson, Hans},
  booktitle = {Proc. of ICTSS},
  title     = {A Survey on Testing for Cyber Physical System},
  year      = {2015},
  pages     = {194--207},
  publisher = {Springer},
  series    = {ICTSS},
  doi       = {10.1007/978-3-319-25945-1_12},
}

@Article{Amann2019,
  author  = {Amann, Sven and Nguyen, Hoan Anh and Nadi, Sarah and Nguyen, Tien N. and Mezini, Mira},
  journal = {Trans. Softw. Eng.},
  title   = {A Systematic Evaluation of Static {API}-Misuse Detectors},
  year    = {2019},
  number  = {12},
  pages   = {1170--1188},
  volume  = {45},
  doi     = {10.1109/TSE.2018.2827384},
}

@InProceedings{Amann2016,
  author    = {Amann, Sven and Nadi, Sarah and Nguyen, Hoan A. and Nguyen, Tien N. and Mezini, Mira},
  booktitle = {Proc. of MSR},
  title     = {{MUBench}: a benchmark for {API}-misuse detectors},
  year      = {2016},
  pages     = {464--467},
  publisher = {ACM},
  series    = {MSR},
  doi       = {10.1145/2901739.2903506},
}

@InProceedings{Bagnara2018,
  author    = {Bagnara, Roberto and Bagnara, Abramo and Hill, Patricia M.},
  booktitle = {Proc. of SAS},
  title     = {The {MISRA} {C} Coding Standard and its Role in the Development and Analysis of Safety- and Security-Critical Embedded Software},
  year      = {2018},
  pages     = {5--23},
  publisher = {Springer},
  series    = {SAS},
  doi       = {10.1007/978-3-319-99725-4_2},
}

@InProceedings{Beyer2017,
  author    = {Beyer, Dirk},
  booktitle = {Proc. of TACAS},
  title     = {Software Verification with Validation of Results (Report on {SV-COMP} 2017)},
  year      = {2017},
  pages     = {331--349},
  publisher = {Springer},
  series    = {TACAS},
  doi       = {10.1007/978-3-662-54580-5_20},
}

@InProceedings{Ball2006,
  author    = {Ball, Thomas and Bounimova, Ella and Cook, Byron and Levin, Vladimir and Lichtenberg, Jakob and McGarvey, Con and Ondrusek, Bohus and Rajamani, Sriram K. and Ustuner, Abdullah},
  booktitle = {Proc. of EuroSys},
  title     = {Thorough static analysis of device drivers},
  year      = {2006},
  pages     = {73--85},
  publisher = {ACM},
  series    = {EuroSys},
  doi       = {10.1145/1217935.1217943},
}

@InProceedings{Clarke2000,
  author    = {Clarke, Edmund and Grumberg, Orna and Jha, Somesh and Lu, Yuan and Veith, Helmut},
  booktitle = {Proc. of CAV},
  title     = {Counterexample-Guided Abstraction Refinement},
  year      = {2000},
  pages     = {154--169},
  publisher = {Springer},
  series    = {CAV},
  doi       = {10.1007/10722167_15},
}

@InCollection{Eichelberger2017,
  author    = {Eichelberger, Hanno and Kropf, Thomas and Ruf, Jürgen and Rosenstiel, Wolfgang},
  booktitle = {Embedded Software Verification and Debugging},
  publisher = {Springer},
  title     = {Automated Reproduction and Analysis of Bugs in Embedded Software},
  year      = {2017},
  editor    = {Lettnin, Djones and Winterholer, Markus},
  pages     = {67--106},
  doi       = {10.1007/978-1-4614-2266-2_4},
}

@InProceedings{Heizmann2018,
  author    = {Heizmann, Matthias and Chen, Yu-Fang and Dietsch, Daniel and Greitschus, Marius and Hoenicke, Jochen and Li, Yong and Nutz, Alexander and Musa, Betim and Schilling, Christian and Schindler, Tanja and Podelski, Andreas},
  booktitle = {Proc. of TACAS},
  title     = {{\ultimateAutomizer} and the Search for Perfect Interpolants (Competition Contribution)},
  year      = {2018},
  pages     = {447--451},
  publisher = {Springer},
  series    = {TACAS},
  doi       = {10.1007/978-3-319-89963-3_30},
}

@Patent{Hill1990,
  nationality = {American},
  number      = {US-4958277},
  year        = {1990},
  yearfiled   = {1989},
  address     = {United States},
  author      = {Susan C. Hill and Joseph Jelemensky and Mark R. Heene and Stanley E. Groves and Daniel N. DeBrito},
  title       = {Queued Serial Peripheral Interface For Use In A Data Processing System},
  type        = {patent},
}

@Article{Henzinger2005,
  author    = {Thomas A. Henzinger and Ranjit Jhala and Rupak Majumdar},
  journal   = {Softw. Eng. Notes},
  title     = {Permissive interfaces},
  year      = {2005},
  number    = {5},
  pages     = {31--40},
  volume    = {30},
  doi       = {10.1145/1095430.1081713},
  publisher = {ACM},
  series    = {FSE},
}

@InCollection{Jhala2018,
  author    = {Jhala, Ranjit and Podelski, Andreas and Rybalchenko, Andrey},
  booktitle = {Handbook of Model Checking},
  publisher = {Springer},
  title     = {Predicate Abstraction for Program Verification},
  year      = {2018},
  editor    = {Edmund M. Clarke and Thomas A. Henzinger and Helmut Veith and Roderick Bloem},
  pages     = {447--491},
  doi       = {10.1007/978-3-319-10575-8_15},
}

@InProceedings{Ko2003,
  author    = {Ko, Amy J. and Myers, Brad A.},
  booktitle = {Proc. of HCC},
  title     = {Development and evaluation of a model of programming errors},
  year      = {2003},
  pages     = {7--14},
  publisher = {IEEE},
  series    = {HCC},
  doi       = {10.1109/HCC.2003.1260196},
}

@InCollection{Lettnin2017,
  author    = {Lettnin, Djones and Winterholer, Markus},
  booktitle = {Embedded Software Verification and Debugging},
  publisher = {Springer},
  title     = {An Overview About Debugging and Verification Techniques for Embedded Software},
  year      = {2017},
  editor    = {Lettnin, Djones and Winterholer, Markus},
  pages     = {1--18},
  doi       = {10.1007/978-1-4614-2266-2_1},
}

@Article{Monperrus2013,
  author    = {Monperrus, Martin and Mezini, Mira},
  journal   = {Trans. Softw. Eng. Methodol.},
  title     = {Detecting Missing Method Calls as Violations of the Majority Rule},
  year      = {2013},
  number    = {1},
  pages     = {1--25},
  volume    = {22},
  doi       = {10.1145/2430536.2430541},
  publisher = {ACM},
}

@TechReport{Motorola2003,
  author = {{Motorola~Inc.}},
  title  = {{SPI} Block Guide},
  year   = {2003},
  number = {S12SPIV3/D},
}

@InProceedings{Nsiah2018,
  author    = {Nsiah, Kofi Atta and Schappacher, Manuel and Rathfelder, Christoph and Sikora, Axel and Groza, Voicu},
  booktitle = {Proc. of I2MTC},
  title     = {An open-source toolkit for retrofit industry 4.0 sensing and monitoring applications},
  year      = {2018},
  pages     = {1--6},
  publisher = {IEEE},
  series    = {I2MTC},
  doi       = {10.1109/I2MTC.2018.8409633},
}

@InCollection{Popovici2009,
  author    = {Popovici, Katalin and Jerraya, Ahmed},
  booktitle = {Hardware-dependent Software},
  publisher = {Springer},
  title     = {Hardware Abstraction Layer},
  year      = {2009},
  editor    = {Ecker, Wolfgang and Müller, Wolfgang and Dömer, Rainer},
  pages     = {67--94},
  doi       = {10.1007/978-1-4020-9436-1_4},
}

@InProceedings{Rathfelder2015,
  author    = {Rathfelder, Christoph and Taspolatoglu, Emre},
  booktitle = {Proc. of SSI},
  title     = {{SensIDL}: Towards a generic framework for implementing sensor communication interfaces},
  year      = {2015},
  pages     = {306--311},
  series    = {SSI},
}

@Article{Strom1986,
  author  = {Strom, Robert E. and Yemini, Shaula},
  journal = {Trans. Softw. Eng.},
  title   = {Typestate: A programming language concept for enhancing software reliability},
  year    = {1986},
  number  = {1},
  pages   = {157--171},
  volume  = {SE-12},
  doi     = {10.1109/TSE.1986.6312929},
}

@InProceedings{Yoo2003,
  author    = {Sungjoo Yoo and Jerraya, A.A.},
  booktitle = {Proc. of DATE},
  title     = {Introduction to hardware abstraction layers for {SoC}},
  year      = {2003},
  pages     = {336--337},
  publisher = {IEEE},
  series    = {DATE},
  doi       = {10.1109/DATE.2003.1253629},
}

@InProceedings{Youssef2004,
  author    = {Youssef, Mohamed-Wassim and Yoo, Sungjoo and Sasongko, Arif and Paviot, Yanick and Jerraya, Ahmed A.},
  booktitle = {Proc. of DAC},
  title     = {Debugging {HW}/{SW} interface for {MPSoC}: video encoder system design case study},
  year      = {2004},
  pages     = {908--913},
  publisher = {ACM},
  series    = {DAC},
  doi       = {10.1145/996566.996808},
}

\end{document}